\documentclass[11pt]{article}
\usepackage{amssymb,dsfont,amsmath,amsfonts,hyperref,mdframed,mathrsfs,stmaryrd}
\usepackage[active]{srcltx}
\usepackage{slashed}
\usepackage{color}
\usepackage{amsthm}
\usepackage{scrtime}
\usepackage{cite}
\usepackage[normalem]{ulem}
\usepackage{framed}
\advance \textheight by \topskip
\newcommand\tsum{\textstyle\sum\nolimits}
\newcommand\tprod{\textstyle\prod\nolimits}

\let\oldsqrt\sqrt
\def\sqrt{\mathpalette\DHLhksqrt}
\def\DHLhksqrt#1#2{%
\setbox0=\hbox{$#1\oldsqrt{#2\,}$}\dimen0=\ht0
\advance\dimen0-0.2\ht0
\setbox2=\hbox{\vrule height\ht0 depth -\dimen0}%
{\box0\lower0.4pt\box2}}

\def\be{\begin{equation}}
\def\ee{\end{equation}}
\def\bea{\begin{eqnarray}}
\def\eea{\end{eqnarray}}

\def\a{\alpha}
\def\b{\beta}

\def\l{\lambda}
\def\s{\sigma}
\def\Comp{\mathbb{C}}
\def\Real{\mathbb{R}}

\input{glyphtounicode}
\begin{document}

\begin{center}
{\Large \textbf{New classes of bi-axially symmetric solutions \\[+5pt]to four-dimensional
Vasiliev higher spin gravity}}\\[0pt]
\vspace*{15mm} Per Sundell\footnote{ per.anders.sundell@gmail.com} and Yihao Yin\footnote{ yinyihao@gmail.com}\\[0pt]
\vspace*{2mm} \textit{Departamento de Ciencias F\'isicas,
  Universidad Andres Bello\\ Republica 220, Santiago de Chile}\\[0pt]
\end{center}

\vspace{1cm}

\begin{minipage}{.90\textwidth}

\paragraph{Abstract}
\bigskip
We present new infinite-dimensional spaces
of bi-axially symmetric asymptotically 
anti-de Sitter solutions to four-dimensional
Vasiliev higher spin gravity, obtained by modifications
of the Ansatz used in arXiv:1107.1217, which gave 
rise to a Type-D solution space.
The current Ansatz is based on internal semigroup
algebras (without identity) generated by exponentials 
formed out of the bi-axial symmetry generators.
After having switched on the vacuum gauge function, 
the resulting generalized Weyl tensor is given by a sum of generalized Petrov type-D tensors that are Kerr-like or 2-brane-like in the asymptotic AdS$_4$ region, and the twistor space connection is smooth in twistor space over finite regions of 
spacetime.
We provide evidence for that the linearized 
twistor space connection can be brought to 
Vasiliev gauge.
\bigskip\bigskip\bigskip
\bigskip\bigskip\bigskip
\bigskip\bigskip\bigskip
\end{minipage}

\vspace{10cm}

\renewcommand{\thefootnote}{\arabic{footnote}}

\setcounter{footnote}{0}

\newpage

{\small \tableofcontents }

\numberwithin{equation}{section}

\section{Introduction}

Vasiliev's equations \cite{Vasiliev:1990en} (for a recent 
review, see \cite{Didenko:2014dwa}) provide 
a fully nonlinear description of higher spin gauge fields 
in four dimensions coupled to gravity and matter fields.
The basic feature of Vasiliev's theory is that the full field 
configurations are captured by master fields that
live on an extension of spacetime by a noncommutative 
twistor space.
The equations admit an exact solution given by the direct
product of anti-de Sitter spacetime and an undeformed
twistor space.
In a specific gauge, certain linearized perturbations of the 
noncommutative twistor space structure give rise to Fronsdal 
fields.
This suggests a holographic relationship to three-dimensional
conformal field theories
\cite{WittenSeminar2001NovJHSchwarz60Birthday,Sezgin:2002rt,Klebanov:2002ja}; 
see also \cite{Sundborg:1999ue,Leigh:2003gk,Koch:2010cy}.
In \cite{Giombi:2009wh,Giombi:2010vg} this relation was examined 
under the assumption that the Gubser-Klebanov-Polyakov-Witten (GKPW) prescription \cite{Gubser:1998bc,Witten:1998qj} for on-shell computations 
of Witten diagrams can be applied to classical field configurations obtained from 
Vasiliev's equations.

However, the Fronsdal fields embedded into Vasiliev's master fields 
have non-local interactions \cite{Sezgin:2002ru,Kristiansson:2003xx,Boulanger:2015ova}
\footnote{For a review, see \cite{Bekaert:2010hw}.} that belong to a functional
class widely separated \cite{Taronna:2016ats} from that 
of the quasi-local Fronsdal theory \cite{Bekaert:2015tva},
which is built by applying the canonical Noether 
approach to Fronsdal fields in anti-de Sitter spacetime.
The GKPW prescription applies to the quasi-local theory 
by construction, as its action has self-adjoint kinetic terms, and 
the resulting holographic correlation functions indeed
correspond to free three-dimensional conformal field 
theories.\footnote{The functional class encountered in the quasi-local Fronsdal theory in \cite{Bekaert:2015tva} 
(within the AdS/CFT context) has not yet been identified completely; for a discussion, see \cite{Taronna:2016ats}, \cite{Bekaert:2016ezc} and Section 7 of \cite{Sleight:2016hyl}. 
At the cubic order, the separation between the functional
classes of this theory and the Vasiliev theory  has been spelled out in 
\cite{Sleight:2016dba}.
}
Recent work \cite{Vasiliev:2016xui} shows that there 
exists an explicit field redefinition that maps 
Vasiliev's theory to a quasi-local theory on-shell, obtained 
by carefully fine-tuning the perturbative expansion on 
the Vasiliev side, though it remains to be seen whether it
coincides with that of \cite{Bekaert:2015tva}.
Moreover, as later shown in \cite{Taronna:2016xrm} the
required field redefinition is large, and hence it is unclear
to what extent the method can be used to actually compute
any holographic correlation functions.
Thus, to our best understanding, the issue of whether  
holographic amplitudes can be extracted by applying
the GKPW prescription to the Fronsdal fields embedded 
into Vasiliev's master fields remains an open problem.

An alternative approach, pursued in \cite{Boulanger:2015kfa}, is to 
seek a weaker relation between the  
two theories, namely at the level of two 
distinct effective actions, derived in their
own rights following different principles, and then evaluated 
subject to suitable dual boundary conditions.
To this end, one starts from Hamilton's principle applied 
to a covariant Hamiltonian action formulated using Weyl order
on a noncommutative manifold whose boundary is given 
by the direct product of spacetime and twistor space 
\cite{Boulanger:2011dd,Boulanger:2015kfa}; the Weyl order
is required for the noncommutative version of the Stokes'
theorem to hold and for the imposition of boundary 
conditions.
The resulting variational principle yields Vasiliev's equations 
in Weyl order, that can be mapped back to Vasiliev's normal 
order for special classes of initial data in twistor space following 
the perturbative scheme set up in \cite{Iazeolla:2011cb,Iazeolla:ToAppear}.
The resulting form of the higher spin amplitudes
\cite{Colombo:2010fu,Colombo:2012jx,Didenko:2012tv}
is closely related to first-quantized topological open string
amplitudes \cite{Engquist:2005yt},
but nonetheless reproduce exactly the same 
correlation functions as the Witten diagrams 
computed in the quasi-local theory.
We would like to stress the fact that the Hamiltonian
form of the action implies that the dependence of the
classical Vasiliev master fields on classical sources are 
of a different type than for fields obeying equations
of motion following from an action with self-adjoint
kinetic terms.
Indeed, instead of applying the GKPW prescription,
the higher spin amplitudes are obtained from functionals
given by topological boundary terms added to the Hamiltonian
action \cite{Sezgin:2011hq,Colombo:2012jx,Boulanger:2015kfa}, 
whose on-shell values are given by higher spin invariants,
as we shall comment on further below.

In this paper, we shall construct new perturbatively defined 
solution spaces to Vasiliev's equations in Weyl order, by taking 
into account classes of functions that resemble closely 
those used in  \cite{Iazeolla:2011cb}.
We shall then demonstrate explicitly that they can be mapped to Vasiliev's
normal order, at least at the linearized level,
thus providing further evidence in favour of the 
covariant Hamiltonian approach outlined
above.

To this end, we recall that at the linearized level, the 
fluctuations in the master fields that are asymptotic 
to anti-de Sitter spacetime form various representation spaces 
of the anti-de Sitter isometry algebra, including lowest-weight
spaces as well as spaces associated to linearized solitons 
\cite{Iazeolla:2008ix} and generalized Petrov type-D solutions 
\cite{Didenko:2009td,Iazeolla:2008ix,Iazeolla:2011cb}.
Nonlinear completions of various Type-D solution 
spaces were constructed in \cite{Didenko:2009td,Iazeolla:2008ix,Iazeolla:2011cb}; for a 
review, see \cite{Iazeolla:2012nf}.
Of direct relevance for the work in this paper is the 
subspace that contains the the black-hole-like solutions,%
\footnote{This subspace is related to the massless spectrum by means of a 
$\mathbf Z_2$-operation \cite{Iazeolla:2011cb}, reminiscent 
of a U-duality transformation \cite{Bossard:2015foa}.}
including spherically symmetric solutions.
In these solutions, each individual Fronsdal field has a 
point-like source at the origin, showing up as a 
divergence in its Weyl tensor.
However, upon packing all curvatures into a master 
zero-form, one obtains the symbol of a quantum-mechanical 
operator that approaches a delta function distribution at 
the origin \cite{Iazeolla:2011cb}, which defines a smooth
state as seen via classical observables given by 
zero-form charges \cite{Sezgin:2005pv,Colombo:2012jx,Didenko:2012tv}.
In this sense, the black-hole-like Type-D solutions to 
Vasiliev's theory are source free at the origin.\footnote{It
remains to be examined whether additional topological two-forms
describing Dirac strings need to be activated in the dynamical
two-form \cite{Vasiliev:2015mka,Boulanger:2015kfa}.}
Furthermore, it is possible to dress these 
solutions with lowest-weight space modes
\cite{Iazeolla:ToAppear} at the fully nonlinear level; in doing so, 
the latter modes induce Type-D modes already at the 
second order of classical perturbation theory.%
\footnote{This phenomena resembles some of the
scattering processes in U-duality covariant field theory 
\cite{Bossard:2015foa}.}

Clearly, the full extent of the moduli space of the theory yet 
remains to be determined.
In this paper, we shall present a new infinite-dimensional 
class of bi-axially symmetric exact solutions that 
are asymptotic to anti-de Sitter spacetime and 
singularity free at the level of zero-form charges.
We shall furthermore propose a super-selection 
mechanism based on requiring that the solutions
can be brought to Vasiliev gauge (where the
asymptotic linearized fluctuations are in terms 
of Fronsdal fields).

Our construction method follows closely the one 
devised in \cite{Iazeolla:2011cb} using gauge functions 
and separation of twistor space variables, which is
in effect equivalent to starting from an Ansatz in Weyl order.
The key difference is that we shall expand the master
fields over a new set of elements in the associative fiber 
algebra, thus adding a branch to the existing moduli space.
In a generic gauge, the expansion coefficients are 
functions on the base manifold.
However, in the holomorphic gauge of  \cite{Iazeolla:2011cb}
the Weyl zero-form is a constant while the twistor 
space one-form is given by a universal set of 
functions, related to Wigner's deformed oscillators,
originally derived within the context of three-dimensional
matter coupled higher spin gravity \cite{Prokushkin:1998bq}.
The resulting solution space is then mapped to Vasiliev gauge
in which the spacetime one-form consists of nonlinear
Fronsdal tensors (after a suitable field redefinition in 
order to reinstate manifest Lorentz covariance).
This map is achieved by means of two consecutive 
(large) gauge transformations: 
First, one uses a vacuum gauge function in 
SO(2,3)/SO(1,3).\footnote{Whether a more
general vacuum gauge function can introduce
additional classical moduli remains an open problem.}
Provided that the resulting twistor space 
connection is smooth at the origin of the base of 
the twistor space, Vasiliev gauge can be 
reached by means of a second perturbatively
defined gauge transformation.
As we shall see, the real-analyticity requirement 
constrains the initial data in the Weyl zero-form
already at the linearized level.%
\footnote{An optional criterion is that the fiber
algebra is a unitarizable representation of the
higher spin algebra and hence the anti-de Sitter 
isometry algebra; we expect this property to arise at higher orders of 
classical perturbation theory by requiring 
positivity of a suitable free energy functional.}

More specifically, the new sector of the fiber algebra 
is isomorphic to the group algebra $\Comp[\mathbb Z\times \mathbb Z]$
where $\mathbb Z\times \mathbb Z$ is generated by two 
elements in Sp$(4;\Comp)$ given by exponentials
of a pair of Cartan generators of sp$(4;\Real)$.
These correspond to linear symmetries of the two-dimensional 
harmonic oscillator, and generate the Killing symmetries of the 
solutions (including higher spin symmetries).
As we shall see, the aforementioned super-selection rule 
amounts to restricting the master fields to a subalgebra of the 
group algebra not containing the unity.

The paper is organized as follows: In Section \ref{Sec Vasiliev eqs}
we review parts of Vasiliev's bosonic higher spin gravity model that
we shall use in constructing and interpreting the exact solutions.
Solution spaces based on (semi)group algebras are constructed 
in Section \ref{Sec solve eqs} using the aforementioned method;
the singular nature of the contribution from the identity
is pointed out in Section \ref{Sec singular id}.
In Section \ref{Sec Weyl tensor}, we show that the Weyl tensor
is given by a sum of Petrov type-D tensors that are Kerr-like or 2-brane-like in the asymptotic AdS$_4$ region, and we compute 
higher spin curvature invariants.
In Section \ref{Sec analyticity}, we show in special cases that the twistor 
space one-form is real-analytic in twistor space over finite regions
of spacetime, and that its linearized part can be brought to
Vasiliev gauge. 
We conclude in Section \ref{Sec Conclusions}.

\section{Bosonic Vasiliev model \label{Sec Vasiliev eqs}}

In this section, we describe the non-minimal bosonic 
higher spin gravity model of Vasiliev type 
\cite{Vasiliev:1990en},\footnote{For recent
reformulations containing the original Vasiliev
system as consistent truncations, see \cite{Vasiliev:2015mka,Boulanger:2015kfa}.}
for which we shall present exact solutions in the
next section.
The model is characterized by the fact that it admits a 
linearization consisting of real Fronsdal fields 
in four-dimensional anti-de Sitter spacetime of spins $s=0,1,2,\dots$ with
each spin occurring once; for further details, we refer 
to \cite{Sezgin:2002ru,Iazeolla:2011cb} and the review \cite{Didenko:2014dwa}.

We first provide the formal definition in terms of master
fields on the direct 
product of a commuting space and
a noncommutative twistor space.
We then spell out the component form of the equations,
including their reformulation in terms of deformed 
oscillators.
Finally, we remark on choices of bases for the
internal algebra, and the Lorentz covariant weak field 
expansion scheme leading to Fronsdal fields, stressing 
the role of Vasiliev gauge and smoothness in twistor 
space.

\subsection{Master field equations}

Vasiliev's original formulation of higher spin gravity 
is given in terms of two master fields $\Phi $ and 
$A$ of degrees $0$ and $1$, respectively, and
two closed and twisted-central elements $I$ and 
$\overline I$ of degree $2$, all of which are elements 
of a differential graded associative algebra $\Omega({\cal M})$ 
of forms on a non-commutative manifold ${\cal M}$, 
valued in an internal associative algebra ${\cal A}$.
Letting $\star$ denote the associative product of 
$\Omega({\cal M})\otimes {\cal A}$, which is assumed to be
compatible with $d$, the fully nonlinear master field equations read
\begin{eqnarray}
F+ {\cal B}\star\Phi \star I- \overline {\cal B}\star\Phi \star  \overline I &=&0 \ ,\label{originalEqA}\\
D\Phi &=& 0\ , \label{originalEqAPhi}\end{eqnarray}
where
\begin{equation}F:=dA+A\star A \ ,\qquad D\Phi:= d\Phi +A\star \Phi -\Phi \star \pi \left( A\right) \ ,\end{equation}
and $\pi$ denotes an automorphism of the differential graded associative algebra.
The two-forms are characterised by the subsidiary constraints
\begin{eqnarray}
dI &=&0\text{ ,} \qquad I\star f \ =\ \pi (f)\star I  \label{Jf=pifJ}\ ,
\end{eqnarray}
for any $f\in \Omega({\cal M})\otimes {\cal A}$, \emph{idem} $\overline I$.
Finally, the star functions
\begin{equation}{\cal B}:=\sum_{n=0}^\infty b_n (\Phi\star \pi(\Phi))^{\star n}\ ,\qquad
\overline {\cal B}:=\sum_{n=0}^\infty \bar b_n (\Phi\star \pi(\Phi))^{\star n}\ ,\end{equation}
where $b_n, \bar b_n\in \Comp$.
It follows that $\Phi\star \pi(\Phi)$ and hence ${\cal B}$ is covariantly constant, \emph{viz.}
\begin{equation}d{\cal B}+A\star {\cal B}-{\cal B}\star A=0\ ,\end{equation}
\emph{idem} $\overline {\cal B}$.
As $\Phi \star I$ and $\Phi \star \overline I$ are covariantly
constant as well, it follows that the constraint on $F$
is compatible with its Bianchi identity.
The integrability of the constraint on $D\Phi$, on the other
hand, requires $F\star \Phi-\Phi\star \pi(F)$ to vanish,
which is indeed a consequence of the constraint on $F$.
The resulting Cartan integrability, \emph{i.e.}
consistency with $d^2\equiv 0$, holds
for any dimension of ${\cal M}$ and any star
functions ${\cal B}$ and $\overline {\cal B}$,
which are hence not fixed uniquely by the requirement
of higher spin symmetry alone.

In the context of higher spin gravity, it is usually
assumed that
\be {\cal M}={\cal X}_4 \times {\cal Z}_4\ ,\label{directproduct}\ee
where ${\cal X}_4$ is a four-dimensional
real commuting manifold, with coordinates $x^{\mu }$, and
${\cal Z}_4$ is a four-dimensional real non-commutative
symplectic manifold, with canonical coordinates
$Z^{\underline{\alpha }}$.
The compatibility between the star product and
the differential amounts to the Leibniz' rule
\begin{equation} d(f\star g)=df\star g+ (-1)^{{\rm deg}(f)} f\star dg\ .\end{equation}
The differential star product algebra is assumed to be trivial 
in strictly positive degrees, in the sense that $d\Xi^M:=(dx^\mu,dz^\a,d\bar z^{\dot{\alpha}})$
are taken to be graded anti-commuting elements obeying
\begin{equation}d\Xi^M \star f= d\Xi^M \wedge f\ ,\qquad
 f\star d\Xi^M = f\wedge d\Xi^M \ ,\end{equation}
which are consistent with associativity.
The algebra $\Omega({\cal M})\otimes {\cal A}$ is also
assumed to be equipped with an anti-linear anti-automorphism
$\dagger$, for which we use the convention
\begin{equation}
\left( f_{1}\star f_{2}\right) ^{\dag }=(-1)^{{\rm deg}(f_1){\rm deg}(f_2)}
f_{2}^{\dag }\star f_{1}^{\dag }\ ,\qquad (df)^\dagger= d(f^\dagger)
\text{ .}
\end{equation}
In case of the basic bosonic models, without internal
Yang-Mills symmetries, the internal algebra ${\cal A}$ consists of 
classes of functions on yet one more four-dimensional 
real non-commutative symplectic manifold, that we shall 
denote by ${\cal Y}_4$, with canonical coordinates 
$Y^{\underline{\alpha}}$.
We shall refer to ${\cal Y}_4\times{\cal Z}_4$ as the 
full twistor space, and ${\cal Y}_4$ and ${\cal Z}_4$, 
respectively, as the internal and external twistor spaces.%
\footnote{Taking the master fields to be smooth 
functions of ${\cal Y}_4$ yields an anti-de Sitter
analog of the Penrose-Newman transformation;
to our best understanding, the precise
relation between ${\cal Y}_4\times{\cal Z}_4$
and the original (commuting) twistor space of Penrose
remains to be spelled out in detail.}
The Sp(4;$\Real$) quartets are split into SL(2;$\mathbb{C}$) 
doublets, \emph{viz.}\footnote{The doublet indices are raised 
and lowered using $f^{\alpha }=\varepsilon ^{\alpha \beta }f_{\beta }$
and $f_{\beta}=f^{\alpha }\varepsilon _{\alpha \beta }$
\emph{idem} $f^{\dot{\alpha} }$. }
\begin{equation} Y^{\underline{\alpha}}=( y^{\alpha },\bar{y}^{\dot{\alpha}})\ ,\qquad
Z^{\underline{\alpha }}= (z^{\alpha },\bar{z}^{\dot{\alpha}})\ ,\end{equation}
obeying
\begin{equation}
\bar{y}^{\dot{\alpha}}=\left( y^{\alpha }\right) ^{\dag }\text{ , }\bar{z}^{
\dot{\alpha}}=-\left( z^{\alpha }\right) ^{\dag }\text{ ,}  \label{realityyz}
\end{equation}
The automorphism $\pi $ and its hermitian conjugate $\bar{\pi}$
are defined by
\begin{eqnarray}
\pi \left( x^{\mu };y^{\alpha },\bar{y}^{\dot{\alpha}};z^{\alpha },\bar{z}^{%
\dot{\alpha}}\right) &=&\left( x^{\mu };-y^{\alpha },\bar{y}^{\dot{\alpha}%
};-z^{\alpha },\bar{z}^{\dot{\alpha}}\right) \text{ ,} \\
\bar{\pi}\left( x^{\mu };y^{\alpha },\bar{y}^{\dot{\alpha}};z^{\alpha },\bar{%
z}^{\dot{\alpha}}\right) &=&\left( x^{\mu };y^{\alpha },-\bar{y}^{\dot{\alpha%
}};z^{\alpha },-\bar{z}^{\dot{\alpha}}\right) \text{ ,}
\end{eqnarray}
and $\pi \circ d=d\circ \pi$ \emph{idem} $\bar \pi$.
Imposing 
\begin{eqnarray}
\Phi ^{\dag } &=&\pi \left( \Phi \right) \text{ ,} \qquad
A^{\dag } \ =\ -A\text{ ,} \qquad I^\dag \ =\ \overline I\ , \label{realityA}
\end{eqnarray}
and 
\begin{equation} {\cal B}^\dag=\overline{\cal B}\ ,\end{equation}
that is, $(b_n)^\dag=\bar b_n$, and
\begin{eqnarray}
\pi \bar{\pi}\left( \Phi \right) &=&\Phi \text{ ,} \qquad  \pi \bar{\pi}\left( A\right) \ =\
A\text{ ,}\qquad \pi \bar{\pi}\left( I \right)  \ =\ I\ ,\qquad \pi \bar{\pi}\left(\overline
I \right)  \ =\ \overline I\label{bosonA}
\end{eqnarray}
yields a model with a perturbative expansion 
around four-dimensional anti-de Sitter
spacetime in terms of Fronsdal
fields of all integer spins.

The equations given so far provide a formal definition of
the basic bosonic model.

\subsection{Star product, twisted central element and traces}\label{starstraces}

In what follows, we shall use Vasiliev's original realization 
of the $\star$-product given by
\begin{eqnarray}
&&f_{1}\left( y,\bar{y},z,\bar{z}\right) \star f_{2}\left( y,\bar{y},z,\bar{z%
}\right)  \notag \\
&=&\!\!\!\!\!\int \frac{d^{2}ud^{2}\bar{u}d^{2}vd^{2}\bar{v}}{\left( 2\pi \right) ^{4}}%
e^{i v^{\alpha }u_{\alpha }+i\bar{v}^{\dot{\alpha}}\bar{u}_{\dot{\alpha}%
} }\ f_{1}\left( y+u,\bar{y}+\bar{u};z+u,\bar{z}-\bar{u}\right)
f_{2}\left( y+v,\bar{y}+\bar{v};z-v,\bar{z}+\bar{v}\right) \text{ .}  \notag
\\
&&  \label{star prod def}
\end{eqnarray}
We shall encounter $\star $-product 
compositions leading to Gaussian integrals involving
indefinite bilinear forms.
To define these we use the fact that the auxiliary 
integration is a formal representation of the original 
Moyal-like contraction formula, which means that the 
integration must be performed by means of analytical 
continuations of the eigenvalues of the bilinear forms.

\paragraph{Symbol calculus.} The star product rule implies that 
\begin{equation}
\left[ f_{1}\left( y,\bar{y}\right) ,f_{2}\left( z,\bar{z}\right) \right]
_{\star }=0\text{ ,}
\end{equation}
that is, the variables $Y^{\underline{\alpha}}$ and $Z^{\underline{\alpha }}$
are mutually commuting.
Moreover, from
\begin{equation}
y_{\alpha }\star y_{\beta }=y_{\alpha }y_{\beta }+i\varepsilon _{\alpha
\beta }\text{ , \ }y_{\alpha }\star z_{\beta }=y_{\alpha }z_{\beta
}-i\varepsilon _{\alpha \beta }\text{ , \ }z_{\alpha }\star y_{\beta
}=z_{\alpha }y_{\beta }+i\varepsilon _{\alpha \beta }\text{ , \ }z_{\alpha
}\star z_{\beta }=z_{\alpha }z_{\beta }-i\varepsilon _{\alpha \beta }\text{ ,%
}  \label{yz comm}
\end{equation}
it follows that
\begin{equation}
a_{\alpha }^{\pm }:=\frac{1}{2}\left( y_{\alpha }\pm z_{\alpha }\right) \text{
,}
\end{equation}
obey
\begin{equation}
\left[ a_{\alpha }^{-},a_{\beta }^{+}\right] _{\star }=
\left[ a_{\alpha }^{+},a_{\beta }^{-}\right] _{\star }=
i\varepsilon _{\alpha \beta }\text{ , }
\left[ a_{\alpha }^{+},a_{\beta }^{+}\right] _{\star }=
\left[ a_{\alpha }^{-},a_{\beta }^{-}\right] _{\star }=0\text{ .}
\end{equation}
Letting ${\cal O}_{\rm Weyl}$ and ${\cal O}_{\rm Normal}$
denote the Wigner maps that send a classical function $f$ to 
the operator with symbol $f$ in the Weyl and normal order,
respectively, where an operator is said to be in normal order 
if all ${\cal O}_{\rm Normal}(a^+_\a)$ stand to the left of 
all ${\cal O}_{\rm Normal}(a^-_\a)$.
As a result, one has
\be {\cal O}_{\rm Normal}(f_{1}\left( y,z\right) \star f_{2}\left(
y,z\right))=
{\cal O}_{\rm Normal}(f_{1}\left( y,z\right)) {\cal O}_{\rm Normal}(f_{2}\left(
y,z\right))
\text{ .}\ee
One also has
\be {\cal O}_{\rm Weyl}(f(y))={\cal O}_{\rm Normal}(f(y))\ ,\qquad
{\cal O}_{\rm Weyl}(f(z))={\cal O}_{\rm Normal}(f(z))\ ,\ee
resulting in that
\begin{eqnarray}
{\cal O}_{\rm Weyl}(f_{1}\left( y\right) \star f_{2}\left( y\right)) &=&
{\cal O}_{\rm Weyl}(f_{1}\left( y\right) ) {\cal O}_{\rm Weyl}( f_{2}\left( y\right)) \text{ ,} \\
{\cal O}_{\rm Weyl}(f_{1}\left( z\right) \star f_{2}\left( z\right)) &=&
{\cal O}_{\rm Weyl}(f_{1}\left( z\right) ) {\cal O}_{\rm Weyl}(f_{2}\left( z\right))\text{ ,}
\end{eqnarray}
and also
\be
{\cal O}_{\rm Normal}( f_{1}\left( y\right) \star f_{2}\left( z\right)) \ =\
{\cal O}_{\rm Weyl}( f_{1}\left( y\right) f_{2}\left( z\right)) \ =\ {\cal O}_{\rm Weyl}( f_{1}\left( y\right) ) {\cal O}_{\rm Weyl}( f_{2}\left( z\right))\text{ .}  \label{f1f2 Weyl}\ee

\paragraph{Twisted central element.}  The condition \eqref{Jf=pifJ} can be solved by
\begin{equation}
I=j_z\star \kappa_y\ ,\qquad j_z=\frac{i}{4}dz^\alpha \wedge dz^\b \varepsilon _{\alpha \beta }\kappa_z 
\text{ , \qquad }\kappa _{y}=2\pi \delta
^{2}\left( y\right) \text{ , \qquad }\kappa _{z}=2\pi \delta ^{2}\left( z\right)
\text{ ,}\label{factorized}
\end{equation}
where $\kappa_y$ is an inner Klein operator obeying
\begin{equation}
\kappa_y \star f(y)\star \kappa_y= f(-y) \text{ , \qquad }\kappa _{y}\star
\kappa _{y}\ =\ 1\ ,
\end{equation}
\emph{idem} $\kappa_z$.
Thus, one may write
\be I=\frac{i}{4}dz^\alpha \wedge dz^\b \varepsilon _{\alpha \beta }\kappa\ ,\qquad
\kappa:=\kappa _{y}\star \kappa _{z}=\exp(i y^\a z_\a)\ ,\label{kappa}\ee
where thus
\be \kappa\star f(y,z)=\kappa f(z,y)\ ,\qquad f(y,z)\star \kappa=\kappa f(-z,-y)\ ,\ee
\be \kappa\star
f(y,z)\star \kappa=\pi(f(y,z))\ ,\qquad \kappa\star\kappa=1\ .\ee
By hermitian conjugation one obtains
\begin{equation}\overline {I}=-I ^{\dag }=%
\frac{i}{4}d\bar z^{\dot\alpha} \wedge d\bar z^{\dot\beta}
\varepsilon _{\dot{\alpha}\dot{\beta}}\bar{\kappa}\ .\end{equation}
The two-forms $j_z$ and $\overline j_{\bar z}$ can be extended to
globally defined forms on a non-commutative space ${\cal Z}_4$ 
having the topology of a direct product of two complexified 
two-spheres \cite{Iazeolla:2011cb,Boulanger:2015kfa}, with
nontrivial flux
\be \int_{{\cal Z}_4} j_z\star \overline j_{\bar z}=-\frac{1}{4}\ . \ee
In this topology, it is furthermore assumed that $\Phi$ belongs
to a section that is bounded at infinity, while the twistor-space 
one-form is a connection whose curvature two-form falls off
at infinity.

We note that the form of $I$ given in Eq.\ \eqref{kappa} is useful 
in deriving the perturbative expansion in terms of Fronsdal 
fields in Vasiliev gauge, while the factorized form in Eq.\ \eqref{factorized} 
is useful in finding exact solutions.

\paragraph{Trace operations.} 
The detailed form of the symbol of an operator depends on the basis 
with respect to which it is defined.
Its trace, on the other hand, is basis independent, and in addition 
gauge invariant. 
The star product algebra admits two natural trace operations.
The basic  operation is given by the integral over 
phase space using the symplectic measure, \emph{viz.}
\be {\rm Tr} f := \int_{{\cal Z}_4\times {\cal Y}_4} 
 j_y\star \overline j_{\bar y} \star \kappa_y \star \bar\kappa_{\bar y}
\star f\ ,\qquad f\in \Omega({\cal Z}_4)\otimes {\cal A}\ ,\ee
where $j_y$ is given by replacing $z^\alpha$ by $y^\alpha$
in $j_z$ defined in Eq.\ \eqref{factorized}.
An alternative trace operation, of relevance to higher spin
gauge theory, can be defined if ${\cal A}$ 
admits the decomposition 
\be {\cal A}=\bigoplus_{n,\bar n=0,1} {\cal A}_{n,\bar n}\star
(\kappa_y)^n \star (\bar\kappa_{\bar y})^{\bar n}\ , \label{calAnbarn}\ee
where ${\cal A}_{n,\bar n}$ consist
of operators whose symbols in Weyl order are 
regular at the origin of ${\cal Y}_4$.
One may then define the trace operation 
\be {\rm Tr}' f := \int_{ {\cal Y}_4} 
j_y\star \overline j_{\bar y} \star f_{1,\bar 1}=-\frac{1}{4} 
 f_{1,\bar 1}|_{y=0=\bar y}\ , \ee
using the decomposition \eqref{calAnbarn}, with the convention that 
\be \kappa_y\star \bar\kappa_{\bar y}\star f=\pm f\quad \Rightarrow\quad 
{\rm Tr}' f =\mp \frac18  f|_{y=0=\bar y}\ .\ee
One may view ${\rm Tr}'$ as a regularized version of ${\rm Tr}$
in the sense that if $f$ admits a decomposition of the form
\eqref{calAnbarn} then
\begin{eqnarray} {\rm Tr} f &=& \sum_{n,\bar n=0,1} {\rm Tr} f_{n,\bar n}\star
(\kappa_y)^n \star (\bar\kappa_{\bar y})^{\bar n}\\
&=&  {\rm Tr}' f+  {\rm Tr} (f_{0,\bar 0} +f_{1,\bar 0}\star
\kappa_y + f_{0,\bar 1} \star \bar\kappa_{\bar y})\ ,
\end{eqnarray}
that is, 
\be {\rm Tr}' f={\rm Tr} f -  {\rm Tr} (f_{0,\bar 0} +f_{1,\bar 0}\star
\kappa_y + f_{0,\bar 1} \star \bar\kappa_{\bar y})\ .
\ee
Indeed, in several applications it turns out that ${\rm Tr} f $
is ill-defined while ${\rm Tr}' f $ is well-defined, as for
example in the case that $f$ is a polynomial on ${\cal Y}_4$.

\subsection{Equations in components and deformed oscillators}

We decompose the master one-form into
locally defined components as follows:
\begin{equation}A=U_{\mu }dx^{\mu }+V_{\alpha }dz^{a}+V_{\dot{\alpha}}d\bar{z}^{\dot{%
\alpha}}\ ,\end{equation}
The reality condition (\ref{realityA}) and the bosonic projection
(\ref{bosonA}) imply
\begin{eqnarray}
U_{\mu }^{\dag } &=&-U_{\mu }\text{ ,} \qquad
V_{\alpha }^{\dag } \ =\ \bar{V}_{\dot{\alpha}}\text{ ,}\\
\pi \bar{\pi}\left( U_{\mu }\right) &=&U_{\mu }\text{ ,} \qquad
\pi \bar{\pi}\left( V_{\alpha }\right) \ =\ -V_{\alpha }\text{ .}
\end{eqnarray}
Decomposing master equations into components using
inner derivatives $\imath_{\partial_\mu}$, $\imath_{\partial_\a}$ and 
$\imath_{\partial_{\dot\a}}$, where $\partial_\a\equiv \partial/\partial z^\a$
\emph{idem} $\partial_{\dot\a}$, one has 
\begin{eqnarray}
\partial _{\lbrack \mu }U_{\nu ]}+U_{[\mu }\star U_{\nu ]} &=&0\text{ ,}
\label{eqxU} \\
\partial _{\mu }\Phi +U_{\mu }\star \Phi -\Phi \star \pi \left( U_{\mu
}\right) &=&0\text{ ,}  \label{eqxPhi}
\end{eqnarray}
the mixed components
\begin{equation}
\partial _{\mu }V_{\alpha }-\partial _{\alpha }U_{\mu }+\left[ U_{\mu
},V_{\alpha }\right] _{\star }=0\text{ , \qquad }\partial _{\mu }\bar{V}_{\dot{
\alpha}}-\partial _{\dot{\alpha}}U_{\mu }+\left[ U_{\mu },\bar{V}_{\dot{
\alpha}}\right] _{\star }=0\text{ ,}  \label{eqxZ}
\end{equation}
which are related by hermitian conjugation, and 
\begin{eqnarray}
\partial _{\lbrack \alpha }V_{\beta ]}+V_{[\alpha }\star V_{\beta ]}+\frac{i%
}{4}\varepsilon _{\alpha \beta }{\cal B}\star \Phi \star \kappa &=&0 \ ,\qquad
\partial_{\lbrack \dot{\alpha}}\bar{V}_{\dot{\beta}]}+\bar{V}_{[\dot{\alpha}}\star
\bar{V}_{\dot{\beta}]}+\frac{i}{4}\varepsilon _{\dot{\alpha}\dot{\beta}}
\overline {\cal B}\star\Phi
\star \bar{\kappa}\ =\ 0\ ,\qquad \label{eqZ-VJ} \\
\partial _{\alpha }\Phi +V_{\alpha }\star \Phi -\Phi \star \bar{\pi}\left(
V_{\alpha }\right) &=&0 \label{eqZ-PhiV} \ ,\qquad
  \partial _{\dot{\alpha}}\Phi +\bar{V}_{
\dot{\alpha}}\star \Phi -\Phi \star \pi \left( \bar{V}_{\dot{\alpha}}\right)
\ =\ 0\text{ ,}\\
\partial _{\alpha }\bar{V}_{\dot{\alpha}}-\partial _{\dot{\alpha}}V_{\alpha
}+\left[ V_{\alpha },\bar{V}_{\dot{\alpha}}\right] _{\star } &=&0\text{ ,}
\label{eqZ-V}
\end{eqnarray}
where the two equations in Eq.\ \eqref{eqZ-VJ} are related by 
hermitian conjugation \emph{idem} Eq.\ \eqref{eqZ-PhiV}.

The twistor space equations \eqref{eqZ-VJ}--\eqref{eqZ-V} can be 
rewritten by introducing Vasiliev's deformed oscillators\cite{Vasiliev:1990en}
\begin{equation}
S_{\alpha }=z_{\alpha }-2iV_{\alpha }\text{ , \ }\bar{S}_{\dot{\alpha}%
}=\bar{z}_{\dot{\alpha}}-2i\bar{V}_{\dot{\alpha}}\text{ ,}
\end{equation}
for which the reality condition and the bosonic projection take the form:
\begin{eqnarray}
\left( S_{\alpha }\right) ^{\dag } &=&-\bar{S}_{\dot{\alpha}}\text{ ,} \\
\pi \bar{\pi}\left( S_{\alpha }\right) &=&-S_{\alpha }\text{ .}
\end{eqnarray}
In terms of the new fields, the aforementioned equations read
\begin{eqnarray}
\left[ S_{\alpha },S_{\beta }\right] _{\star } &=&-2i\varepsilon _{\alpha
\beta }\left( 1-{\cal B}\star \Phi \star \kappa \right) \text{ \ and h.c.\ ,} \\
S_{\alpha }\star \Phi +\Phi \star \pi \left( S_{\alpha }\right) &=&0\text{ \
and h.c.\ ,} \\
\left[ S_{\alpha },\bar{S}_{\beta }\right] _{\star } &=&0\text{ ,}
\end{eqnarray}
as can be seen using 
\begin{eqnarray}
\left[ z_{\alpha },f\right] _{\star } &=&-2i\partial _{\alpha }f\text{ , \ }%
\left[ \bar{z}_{\dot{\alpha}},f\right] _{\star }=-2i\partial _{\dot{\alpha}}f%
\text{ ,} \\
\left[ z_{\alpha },z_{\beta }\right] _{\star } &=&-2i\varepsilon _{\alpha
\beta }\text{ , \ }\left[ \bar{z}_{\dot{\alpha}},\bar{z}_{\dot{\beta}}\right]
_{\star }=-2i\varepsilon _{\dot{\alpha}\dot{\beta}}\text{ ,} \\
\left[ z_{\alpha },\bar{z}_{\dot{\alpha}}\right] _{\star } &=&0\text{ .}
\end{eqnarray}
As we shall see below, the deformed oscillators are useful in
defining the field redefinition to Lorentz covariant basis.
They also provide a useful basis for finding exact solutions
as they convert the differential equations on ${\cal Z}_4$ into
algebraic equations that can be solved using Laplace
transformation methods \cite{Prokushkin:1998bq}; for related details, 
see \cite{Iazeolla:2011cb}.

\subsection{Lorentz covariance, Fronsdal fields and Weyl tensors}
\label{LorentzFronsdal}

To arrive at a perturbative formulation in terms of Fronsdal fields
on ${\cal X}_4$, one first solves Eqs.\ (\ref{eqxZ})--(\ref{eqZ-V})
subject to an initial datum for $\Phi$ and $U_\mu$ at 
$Z^{\underline\a}=0$ in a perturbative expansion 
in the zero-form initial data in Vasiliev gauge%
\footnote{At the linearized level, this gauge yields the
canonical basis for unfolded linearized Fronsdal
fields \cite{Vasiliev:1990en}; for further details,
see \cite{Sezgin:2002ru} and the review \cite{Didenko:2014dwa}.
Beyond the linearized approximation, it has
been used in amplitude computations
\cite{Giombi:2009wh,Giombi:2010vg,Colombo:2010fu,Colombo:2012jx}
and related recent works\cite{Vasiliev:2015mka,Vasiliev:2016sgg}.
Most exact solutions found so far, however, have been
given in other gauges argued to be equivalent
to Vasiliev gauge; for example, see
\cite{Sezgin:2005pv,Didenko:2008va,Iazeolla:2011cb}.}
\begin{equation}z^\alpha V_\alpha=0\ .\label{vasgauge}\end{equation}
In this gauge, initial data for the zero-form given by
generic smooth symbols on ${\cal Y}_4$
yields twistor space configurations that are 
smooth functions on ${\cal Y}_4\times {\cal Z}_4$.
Letting $\omega_\mu^{\alpha\beta}$ denote the 
canonical Lorentz connection, one can show that 
\cite{Vasiliev:1999ba} $\Phi$, $V_\a$ and%
\footnote{The resulting manifestly Lorentz covariant form of the
master field equations can be found in \cite{Iazeolla:2011cb,Colombo:2010fu}.}
\begin{equation} W_\mu:=U_\mu- \frac1{4i} \left( \omega_\mu^{\alpha\beta} {M}_{\alpha\beta}
+\bar{\omega}_\mu^{\dot{\alpha}\dot{\beta}} M_{\dot{\alpha}\dot{\beta}}\right)\ ,
\label{fieldredef}\end{equation}
where
\begin{equation} M_{\alpha\beta}:=y_\alpha y_\beta -z_\alpha z_\beta+S_{\alpha}
 \star S_{\beta}\ ,\end{equation}
have Taylor expansions in $(Y^{\underline\a},Z^{\underline\a})$
around $Y^{\underline\a}=Z^{\underline\a}=0$ in terms of Lorentz
tensors.
The redefinition induces a shift symmetry that can be used to
set the coefficient of $y_\alpha y_\beta$ in $W_\mu$ to zero,
such that
\begin{equation}W_\mu|_{Z=0}=e_\mu+W'_\mu\ ,\qquad e_\mu=\frac1{2i}
e^{\alpha\dot\alpha}_\mu y_\alpha \bar y_{\dot \alpha}\ ,\end{equation}
where $W'_\mu$ consists of a spin-one field and
a tower of higher spin gauge fields with $s=3,4,\dots$.
Proceeding by assuming that $e^{\alpha\dot\alpha}_\mu$
defines a vierbein, and taking $\Phi|_{Z=0}$ and  $W'_\mu$ 
to be weak fields in which the couplings in Eqs.\ (\ref{eqxU})--(\ref{eqxPhi}) 
can be expanded perturbatively, one can show that
the resulting algebraically independent fields are 
given by the Lorentz scalar
\begin{equation}\varphi:=\Phi|_{Y=Z=0}\ ,\label{Vasilievgauge}\end{equation}
the metric
\begin{equation} 
g_{\mu\nu}:=e_\mu^a e_{\nu,a}\ ,
\end{equation}
and the tower of doubly traceless tensor gauge fields
\begin{equation} \varphi_{a_1\dots a_s}:=(e^{-1})_{(a_1}{}^\mu
W'_{\mu,a_2\dots a_s)}\ ,\qquad s=1,3,4,\dots\ ,\label{Fronsdal}
\end{equation}
where $W'_{\mu,a_1\dots a_n}$ is the coefficient in 
$W'_\mu$ of $(\sigma^{a_1})_{\a\dot\a}y^{\a}\bar y^{\dot{\a}}\cdots 
(\sigma^{a_n})_{\a\dot\a}y^{\a}\bar y^{\dot{\a}}$.
These fields obey equations of motion on the Lorentzian 
manifold $({\cal X}_4,g_{\mu\nu})$ with second-order 
kinetic terms, critical masses and dynamical metric.%
\footnote{Whether the resulting system admit any consistent truncation to a
pure higher-derivative gravity theory remains an open problem.}

The virtue of Vasiliev gauge is that the metric
and the gauge fields \eqref{Fronsdal} are identical to 
the Fronsdal tensors that can be obtained at the
linearized level by integrating the generalized Weyl 
tensor
\be C_{\a_1\dots \a_{2s}}=\left.\left(\frac{\partial^{2s}}{\partial y^{\a_1}\cdots
\partial y^{\a_{2s}}} \Phi\right)\right|_{Y=Z=0}\ ,\qquad s=1,2,3,\dots\ ,\ee
using the generalized Poincare lemma (for example, see \cite{DuboisViolette:1999rd,DuboisViolette:2001jk,Bekaert:2002dt}).
In other words, an asymptotic observer who sources the bulk
using a linearized spin-$s$ Fronsdal field will activate the 
corresponding component field given above, whose boundary
value can thus be identified with a dual conformal field theory 
source coupled to a conserved spin-$s$ current.

The higher order couplings depend on the choice
of gauge as well as the initial data for $\Phi$ and $W_\mu$;
as proposed by Vasiliev \cite{Vasiliev:2016xui},
these initial data can be fine-tuned at higher orders
in order to obtain quasi-local equations of motion
in the gauge \eqref{vasgauge}.

An alternative approach, which we shall follow here,
is to restrict the initial data for the zero-form to 
specific classes of functions on ${\cal Y}_4$,
corresponding to associative subalgebras of ${\cal A}$
leading to well-defined field configurations
obeying physical boundary conditions on ${\cal M}$.

\subsection{Internal star product algebras and solution spaces}

A parameterised set $(\Phi(\nu,G),U(\nu,G),V(\nu,G),
\bar V(\nu,L))_{}$, where $\nu$ belongs
to a parameter space and $G$ is a gauge function, 
obeying the master field equations form an admissible 
solution space if they generate a free 
differential algebra together with $I$ and $\overline I$ 
(for each fixed value of $\nu$).
To construct such spaces we use associative star product 
algebras\footnote{The multiplication 
	table of ${\cal A}_{\cal S}$
	may involve fusion rules \cite{Iazeolla:2011cb,Boulanger:2013naa}, 
	which stipulate which pairs of basis elements that have 
	nontrivial star products and which basis elements that
	are to be used to expand the result.} 
\be {\cal A}_{\cal S}=\bigoplus_{\lambda\in {\cal S}} T_\l \otimes \Comp\ ,\ee
that are closed under the actions of $\pi$, $\bar\pi$, $\dagger$ and star multiplication 
by $\kappa_y$ and $\bar\kappa_{\bar y}$, and whose basis elements $T_\l$, labeled by $\lambda$ 
in a discrete set ${\cal S}$, have finite traces. 
We say that ${\cal A}_{{\cal S}}$ is contained 
in ${\cal A}_{{\cal S}'}$ if there exists
a monomorphism $\rho:{\cal A}_{{\cal S}'}\to 
{\cal A}_{{\cal S}}$ such that ${\rm Tr}'\circ \rho={\rm Tr}'$ 
\emph{i.e.} if the elements in ${\cal A}_{{\cal S}}$ can be 
expanded in terms of the elements in ${\cal A}_{{\cal S}'}$
in a way compatible with the trace operation.

Expanding the master fields over ${\cal A}_{\cal S}$ 
yields a set of modes on ${\cal X}_4$ and ${\cal Z}_4$ that
forms a free differential algebra together with 
$j_z$ and its hermitian conjugate.
Using Cartan integration methods, the modes can be expressed
locally in terms of zero-form integration constants, which
define the $\nu$ parameters, and gauge functions.
These data can then be adapted to boundary conditions,
which may require a change of basis from ${\cal A}_{{\cal S}}$
to a basis ${\cal A}_{{\cal S}'}$ containing ${\cal A}_{{\cal S}}$; for example, in asymptotically 
anti-de Sitter spacetimes, it makes sense to impose boundary conditions   
in a Lorentz covariant basis adapted to a dual conformal field theory.
We shall say that a subalgebra ${\cal A}_{\cal S}$ yields a 
higher spin gravity solution space if the resulting Lorentz covariant master 
fields in Vasiliev gauge have symbols defined in
normal order that can be expanded over finite regions 
of ${\cal X}_4$ in terms of the set 
of monomials on ${\cal Y}_4\times {\cal Z}_4$ that vanish 
at the origin of ${\cal Y}_4\times {\cal Z}_4$, \emph{i.e.}
they are real-analytic at this point. 

The resulting moduli spaces can be coordinatized by 
higher spin invariant functionals, playing the role of classical 
higher spin observables \cite{Colombo:2010fu,Sezgin:2011hq,Vasiliev:2015mka,Vasiliev:2016sgg}.
By choosing a structure group \cite{Sezgin:2011hq} and fixing a 
topology for the base manifold, one may extend the locally
defined solutions to globally defined higher spin geometries
supporting various types of topologically nontrivial observables.
Working locally on ${\cal X}_4$, the accessible observables are 
on-shell closed zero-forms on ${\cal X}_4$ given by 
combined integrals over ${\cal Z}_4$ and traces over
${\cal Y}_4$ of adjoint constructs built from 
$(\Phi,V_\a,\bar V_{\dot\a};I,\overline I;\kappa,\bar\kappa)$,
referred to as zero-form charges.
Evaluated on solutions that are asymptotical
to anti-de Sitter spacetime, these observables 
have been shown to have a physical interpretation
as generating functionals for correlation functions
of holographically dual conformal field theories.

We remark that various subalgebras  
of ${\cal A}$ can be obtained from different quantum mechanical 
systems in four-dimensional phase space.
It is an interesting problem to examine which
of these are admissible in the above sense,
and to furthermore distinguish between these systems
using higher spin invariant observables.
%

\section{New class of biaxially symmetric solutions\label{Sec solve eqs}}

In this section, we construct a new class of exact
solutions to Vasiliev's equations on a 
direct product manifold of the form \eqref{directproduct}
using a gauge function and expansions in terms of exponentials 
of two Cartan generators of ${\rm sp}(4;\Comp)$, which leads
to biaxial symmetry.

\subsection{Gauge function}\label{Sec Gauge function}

From Eq.\ (\ref{eqxU}) and the fact that ${\cal X}_4$ is commuting, 
it follows that $U_{\mu}$ can be expressed in terms of a gauge 
function $G$ defined locally on ${\cal X}_4\times {\cal Z}_4$ .
Thus, setting\footnote{We denote the star product
inverse of $G$ by $G^{-1}$, that is, $G\star G^{-1}=1$.}
\begin{eqnarray}
U^{(G)}_{\mu } &=&G^{-1}\star \partial _{\mu }G\text{ ,}  \label{AnsatzUG} \\
\Phi^{(G)} &=&G^{-1}\star \Phi ^{\prime }\star \pi {}\left( G\right)
\label{AnsatzPhiG} \\
V^{(G)}_{\alpha } &=&G^{-1}\star \partial _{\alpha }G+G^{-1}\star V_{\alpha
}^{\prime }\star G\text{ , }\qquad\bar{V}^{(G)}_{\dot{\alpha}}=G^{-1}\star \partial _{%
\dot{\alpha}}G+G^{-1}\star \bar{V}_{\alpha }^{\prime }\star G \text{ ,}
\label{AnsatzVG}
\end{eqnarray}
Eqs.\ (\ref{eqxPhi}) and (\ref{eqxZ}) reduce to 
\begin{equation} \partial_\mu  \Phi ^{\prime }=0\ ,\qquad \partial_\mu V_{\alpha
}^{\prime }=0\ ,\qquad \partial_\mu \bar{V}_{\alpha }^{\prime }=0\ ,\end{equation}
\emph{i.e.} the primed fields are constant on ${\cal X}_4$, and Eqs.\ (\ref{eqZ-VJ})-(\ref{eqZ-V}) take the form
\begin{eqnarray}
\partial _{\lbrack \alpha }V_{\beta ]}^{\prime }+V_{[\alpha }^{\prime }\star
V_{\beta ]}^{\prime }+\frac{i}{4}\varepsilon _{\alpha \beta }{\cal B}'\star \Phi ^{\prime
}\star \kappa &=&0\text{ \ and h.c.\ ,}  \label{X-indep-VJ} \\
\partial _{\alpha }\Phi ^{\prime }+V_{\alpha }^{\prime }\star \Phi ^{\prime
}-\Phi ^{\prime }\star \bar{\pi}\left( V_{\alpha }^{\prime }\right) &=&0%
\text{ \ and h.c.\ ,}  \label{X-indep-PhiV} \\
\partial _{\alpha }\bar{V}_{\dot{\alpha}}^{\prime }-\partial _{\dot{\alpha}%
}V_{\alpha }^{\prime }+\left[ V_{\alpha }^{\prime },\bar{V}_{\dot{\alpha}%
}^{\prime }\right] _{\star } &=&0\text{ ,}  \label{X-indep-V}
\end{eqnarray}
where ${\cal B}':=\sum_{n=0}^\infty b_n (\Phi'\star \pi(\Phi'))^{\star n}$.

In order to obtain solutions that are asymptotic 
to AdS$_4$, we choose%
\footnote{The reality condition and bosonic projection of
a gauge function $G$ takes the form $G^{\dag }=G^{-1}$
and $\pi \bar{\pi}\left( G\right)= G$.}
\begin{equation} G=L\star H\ ,\end{equation}
where $L$, which we shall refer to as the vacuum
gauge function, is a locally defined map from ${\cal X}_4$ to 
SO(2,3)/SO(1,3) that is constant on ${\cal Z}_4$, \emph{i.e.}
\begin{equation}\partial _{\alpha}L=\partial _{\dot{\alpha}}L=0\ ,
\end{equation}
and $H$ is determined by imposing the Vasiliev gauge
condition \eqref{vasgauge}, \emph{viz.}
\begin{equation} z^\a V^{(G)}_{\alpha } =0\ ,\qquad
\bar z^{\dot\alpha} \bar{V}^{(G)}_{\dot{\alpha}}=0\ ,\end{equation}
in a perturbative expansion
\begin{equation} H=1+\sum_{n=1}^\infty H^{(n)}\ ,\end{equation}
where the superscript $(n)$ denotes an $n$-linear function of $\Phi'$.
Thus, the master fields in Vasiliev gauge are given by
perturbative corrections of 
\begin{eqnarray}
U^{(L)}_{\mu } &=&L^{-1}\star \partial _{\mu }L\text{ ,}  \label{AnsatzU} \\
\Phi^{(L)} &=&L^{-1}\star \Phi ^{\prime }\star \pi {}\left( L\right)
\label{AnsatzPhi} \\
V^{(L)}_{\alpha } &=&L^{-1}\star \partial _{\alpha }L+L^{-1}\star V_{\alpha
}^{\prime }\star L\text{ , }\qquad\bar{V}^{(L)}_{\dot{\alpha}}=L^{-1}\star \partial _{%
\dot{\alpha}}L+L^{-1}\star \bar{V}_{\alpha }^{\prime }\star L \text{ ,}
\label{AnsatzV}
\end{eqnarray}
where the Maurer-Cartan form $U^{(L)}_{\mu}$ 
consists of the frame field and Lorentz connection
on the anti-de Sitter background spacetime, for which
we shall use the explicit form in stereographic coordinates 
given in Appendix \ref{Sec coordinates}.
As for $H$, its existence requires that $V^{(L)}_{\underline{\alpha}}$ 
admits a power series expansion on ${\cal Z}_4$ around 
$Z^{\underline{\alpha}}=0$, to be examined in more detail 
in Section \ref{Sec analyticity}.

Thus, the dependence on ${\cal X}_4$ arises via the gauge function,
leaving ${\cal X}_4$-independent equations (\ref{X-indep-VJ})--(\ref{X-indep-V}),
to which we turn next.

\subsection{Exact solutions in holomorphic gauge from abelian group algebras} \label{Sec general ansatz}

One class of solution spaces arise from star product algebras 
\be {\cal A}_{\Lambda}=\bigoplus_{n,\bar n=0,1}\bigoplus_{\vec\lambda\in \Lambda} 
\left(T_{\vec\l}
\star\kappa_y^{\ n}\star  \bar\kappa_{\bar y}^{\ \bar{n}}\right) \otimes \Comp
\ ,\label{calALambda}\ee
where $\vec\l=(\l_1,\dots,\l_N)$ belongs to an $N$-dimensional lattice $\Lambda$ and 
\be T_{\vec \l}\star T_{\vec \l'}=T_{\vec \l+\vec\l'}\ ,\qquad [T_{\vec \l},  \kappa _{y}\star \bar{\kappa}_{
\bar{y}}]_\star =0\ ,\qquad ( T_{\vec \l})^\dagger =  T_{c(\vec \l)}\ ,\qquad \pi(T_{\vec \l})=T_{\pi(\vec \l)}\ ,
\label{basis}\ee
for $c,\pi:\Lambda\to \Lambda$. 
The second relation, which is equivalent to the bosonic projection 
$\pi \bar{\pi}\left( T_{\vec\l}\right) =T_{\vec\l}$, makes it possible to 
decompose under
\begin{equation}
\Pi _{\sigma }:=\frac{1}{2}\left( 1+\sigma \kappa _{y}\star \bar{\kappa}_{
\bar{y}}\right) =\frac{1}{2}\left( 1+\sigma \kappa _{y}\bar{\kappa}_{\bar{y}
}\right) \text{ ,}  \label{Pidef}
\end{equation}
by expanding 
\begin{eqnarray}
\Phi ^{\prime } &=&\sum_{\sigma; \vec\l}T_{\vec\l}
\star \Pi _{\sigma }\star ( \nu _{\sigma; \vec\l} \kappa _{y}+ \check
\nu _{\sigma; \vec\l})\text{ ,} \qquad \ \nu _{\sigma;\vec\l}\ , \ \check\nu _{\sigma;\vec\l}\in \mathbb{C}\ , \label{Phiprime}\\
V_{\alpha }^{\prime } &=&\sum_{\sigma;\vec\l} T_{\vec\l}\star \Pi
_{\sigma }\star \left( a_{\sigma;\vec\l;\a}+ \check a_{\sigma;\vec\l;\a}\star \kappa _{y}\right)\text{ ,}
\label{Vprime}
\end{eqnarray}
where $ a_{\sigma;\vec\l;\a}$ and $\check a_{\sigma;\vec\l;\a}$ are
holomorphic functions on ${\cal Z}_4$ and are constant over ${\cal Y}_4$, which may be viewed
as a gauge choice (for given zero-form initial data).
Expanding\footnote{We use a convention such that if ${\cal B}'=b_0$ 
then $\mu _{\sigma; \vec\l}=b_0 \nu _{\sigma; \vec\l}$
and $\check
\mu _{\sigma; \vec\l}=b_0 \check
\nu _{\sigma; \vec\l}$.}
\be {\cal B}'\star \Phi'\star \kappa_y=\sum_{\sigma; \vec\l}T_{\vec\l}
\star \Pi _{\sigma }\star ( \mu _{\sigma; \vec\l} + \check
\mu _{\sigma; \vec\l} \,\kappa _{y})\text{ ,} \qquad \ \mu _{\sigma;\vec\l}\ , \ \check\mu _{\sigma;\vec\l}\in \mathbb{C}\ ,\label{defmu}\ee
and introducing 
\begin{equation}
\mathring{\mu}_{\sigma }( \vec\zeta) :=\sum_{\vec\l}\mu
_{\sigma;\vec\l}(\vec\zeta)^{\vec\l}\ ,\qquad
\mathring{\check\mu}_{\sigma }( \vec\zeta) :=\sum_{\vec\l}\check\mu
_{\sigma;\vec\l}(\vec\zeta)^{\vec\l}\ ,\ee
\be 
\mathring{a}_{\sigma }(\vec\zeta):= \sum_{\vec\l} dz^\a {a}_{\sigma;\vec\l;\alpha }
(\vec\zeta)^{\vec\l}\ ,
\qquad
\mathring{\check a}_{\sigma }(\vec\zeta):= \sum_{\vec\l} dz^\a {\check a}_{\sigma;\vec\l;\alpha }
(\vec\zeta)^{\vec\l}\ ,
\end{equation}
where $\vec\zeta:=(\zeta_1,\dots,\zeta_N)\in \Comp^N$ and 
$(\vec\zeta)^{\vec\l}:=(\zeta_1)^{\l_1}\cdots(\zeta_N)^{\l_N}$,
the remaining equations on ${\cal Z}_4$ take the form
\be d\mathring{a}_{\sigma }+\mathring{a}_{\sigma }\star
\mathring{a}_{\sigma }+\mathring{\check a}_{\sigma }\star \gamma \star
\mathring{\check a}_{\sigma }\star\gamma+ j_z \mathring{\mu}_{\sigma }=0\ ,
\label{3.23}\ee
\be d\mathring{\check a}_{\sigma }\star \gamma +\mathring{a}_{\sigma }\star
\mathring{\check a}_{\sigma }\star \gamma +\mathring{\check a}_{\sigma }\star \gamma \star
\mathring{a}_{\sigma }+ j_z \mathring{\check \mu}_{\sigma }\star \gamma=0\ ,\ee
where the element $\gamma$ obeys
\be \gamma\star (\vec\zeta)^{\vec\l}\star \gamma=(\vec\zeta)^{\pi(\vec\l)}\ ,\qquad [\gamma,z^\a]_\star=0\ .\ee
Defining 
\be\mathring{\mu}_{\sigma}^\pm=\mathring{\mu}_{\sigma}\pm
\mathring{\check \mu}_{\sigma}\star\gamma\ ,\qquad 
\mathring{a}_{\sigma}^\pm = \mathring{a}_{\sigma }\pm \mathring{\check a}_{\sigma }\star \gamma\ ,\label{defamu}\ee
we obtain two decoupled systems of the form
\be d\mathring{a}_{\sigma}^\pm +\mathring{a}_{\sigma}^\pm \star
\mathring{a}_{\sigma}^\pm + j_z \mathring{\mu}_{\sigma}^\pm =0\ ,\label{3.27}\ee
that can be solved using the method of \cite{Iazeolla:2011cb}
(see also \cite{Iazeolla:2012nf}), drawn from the
original method devised in \cite{Prokushkin:1998bq}.
Omitting discrete moduli which arise via projector algebras on ${\cal Z}_4$,
two particular solutions that we label by $\varsigma =\pm 1$,
are given by
\begin{equation}
\left( \mathring{a}_{\sigma;\varsigma }^\pm \right) _{\alpha }=2iz_{\alpha
}\int_{-1}^{1}\frac{d\tau  }{\left( \tau
+1\right) ^{2}} j_{\sigma }\left( \varsigma \mathring{\mu}
_{\sigma}^\pm \ ;\ \tau \right)\text{exp}\left[ \varsigma c\left( \tau \right) U^{\beta \gamma
}z_{\beta }z_{\gamma }\right] \text{ ,}  \label{a-circle soln}
\end{equation}
where
\be
j_{\sigma }\left(\varsigma \mathring{\mu}_{\sigma}^\pm;\tau \right) :=-
\frac{\varsigma \mathring{\mu}_{\sigma}^\pm}{4}{}_{1}F_{1}\left[ \frac{1}{2};2;\frac{\varsigma \mathring{\mu}_{\sigma}^\pm}{2}\text{log} \,\tau ^{2} \right] \text{ ,}
\qquad c\left( \tau \right) := i\frac{\tau -1}{\tau +1}\text{ , } \ee
and
\begin{equation}
U^{\beta \gamma }:=\left( u^{+}\right) ^{(\beta }\left( u^{-}\right)
^{\gamma )}\text{,}
\end{equation}
where $u^{+}$ and $u^{-}$ are a set of spinor basis vectors obeying
\begin{equation}
\left( u^{+}\right) ^{\alpha }\left( u^{-}\right) _{\alpha }=1\text{ , \ }%
\left( u^{+}\right) ^{\alpha }\left( u^{+}\right) _{\alpha }=\left(
u^{-}\right) ^{\alpha }\left( u^{-}\right) _{\alpha }=0\text{ .}
\end{equation}
Using (\ref{eps explicit}), we can choose
\begin{equation}
\left( u^{+}\right) ^{\alpha }=\left[
\begin{array}{c}
0 \\
1%
\end{array}%
\right] \text{ , \ }\left( u^{-}\right) ^{\alpha }=\left[
\begin{array}{c}
1 \\
0%
\end{array}%
\right] \text{ .}
\end{equation}
The original twistor space connection can thus be obtained
by expanding the confluent hypergeometric function in a 
power series, followed by identifying powers of $\vec\zeta$ and
$\gamma$, though in what follows we shall mainly work directly
with the generating functions.

\subsection{Twistor space connection in Weyl order in holomorphic gauge}
\label{Singular Integrand}

The twistor space connection $V'_\a$ is given in the 
holomorphic gauge by \eqref{Vprime}.
From Eq.\ \eqref{f1f2 Weyl}, it follows that
\begin{eqnarray}
&&{\cal O}_{\rm Normal}\left(\sum_{\sigma;\vec\l} T_{\vec\l}\star \Pi
_{\sigma }\star \left( a_{\sigma;\vec\l;\a}+ \check a_{\sigma;\vec\l;\a}\star \kappa _{y}\right)\right)
\notag \\
&=& {\cal O}_{\rm Weyl}\left(\sum_{\sigma;\vec\l} \left( (T_{\vec\l}\star \Pi
_{\sigma })  a_{\sigma;\vec\l;\a} + (T_{\vec\l}\star \Pi
_{\sigma }\star \kappa _{y}) \check a_{\sigma;\vec\l;\a}\right)\right)\ ,
\end{eqnarray}
that is, the symbol in Weyl order of $V'_\a$ is given by the argument
of the Wigner map on the right-hand side. 
This quantity contains singular distributions on ${\cal Y}_4$,
which we shall examine in more detail later, and on ${\cal Z}_4$,
which we shall examine in what follows.
To this end, we observe that the integrand in (\ref{a-circle soln}) 
has potential divergences at $\tau=0$, where log($\tau^2$) goes to infinity,
and at $\tau=-1$, where denominators vanish.

As for the potential divergence at $\tau=0$, it does not lead to any 
non-real-analyticity in ${\cal Z}_4$ to any finite
order in perturbation theory as follows from 
the fact that%
\footnote{The confluent hypergeometric function 
${}_1F_1(a;b;x):=\sum_{n=0}^\infty \frac{(a)_n x^n }{(b)_n n!}$ 
obeys $0<{}_1F_1(a;b;x)< e^x$ for $b>a>0$ and $x>0$.
Its asymptotic form for large $|x|$ is given by ${}_1F_1(a;b;x)\sim
\frac{\Gamma(b)}{ \Gamma(a)} x^{a-b} e^x +
\frac{\Gamma(b)}{ \Gamma(b-a)} (-x)^{-a}$.}
\begin{equation}
\left\vert {}_{1}F_{1}
\left[ \frac{1}{2};2;\frac{\varsigma \mathring{\mu}^\pm_\sigma}{2}\log
\left( \tau ^{2}\right) \right] \right\vert \leq \left\vert {}\tau
^{\varsigma \mathring {\mu}^\pm_\sigma}\right\vert\ ,\qquad
{\rm Re} (\varsigma \mathring{\mu}^\pm_\sigma) <0\ ,\
\end{equation}
for $\tau\in [-1,1]$, while the same quantity is bounded 
for $\tau\in [-1,1]$ if ${\rm Re} (\varsigma \mathring{\mu}^\pm_\sigma) \geqslant 0$.
Thus, at $\tau=0$ there is no singularity as long as $\mathring{\mu}^\pm_\sigma$
lies inside the unit disc; indeed, for $\mathring{\mu}^\pm_\sigma$ 
sufficiently close to zero, the power series expansion of the confluent hypergeometric function yields a basis of functions of $\tau$ that can be used to convert the integral equation, 
obtained by inserting Eq.\ (\ref{a-circle soln}) into the deformed
oscillator equation, into an algebraic equation for symbols (for details, see \cite{Prokushkin:1998bq,Iazeolla:2011cb}).
Thus, in order for (\ref{a-circle soln}) to provide a solution, there has to
exist an annulus of convergence in the $\vec \zeta$-space
for the Laurent series defining $\mathring{\mu}^\pm_\sigma$ where 
its modulus is less than one, which can be achieved by tuning 
the overall strength of the $\nu$- and $b_n$-parameters.
In other words, the contribution to (\ref{a-circle soln}) from the
region around $\tau=0$ is real-analytic on ${\cal Z}_4$ to any
finite order in perturbation theory.

Turning to the divergence at $\tau=-1$, it induces a simple pole
$\mathring a^\pm_\a|_{\rm pole}$ in $\mathring a^\pm_\a$ at $z^\a=0$,
which can be extracted using the formula
\be \int_{-1}^1 \frac{d\tau}{(\tau+1)^2} e^{\frac{\tau-1}{\tau+1}p }=\frac{1}{2p}\ ,\qquad
{\rm Re}\, p>0\ ,\ee
and analytical continuation of $U^{\beta \gamma }z_{\beta
}z_{\gamma }$.
It follows that 
\begin{eqnarray}
(\mathring a^\pm_{\sigma;\varsigma})_\a|_{\rm pole}&=&\left.2iz_{\alpha }\int_{-1}^{1}
\frac{d\tau}{\left( \tau +1\right) ^{2}} j_{\sigma }\left( \varsigma \mathring{\mu}^\pm
_{\sigma };\tau \right)\text{exp}\left[ 
\frac{\varsigma i(\tau-1)}{\tau +1}U^{\beta \gamma }z_{\beta }z_{\gamma }\right] 
\right|_{\rm pole}
\notag \\
&=&\left.-iz_{\alpha }\frac{\varsigma \mathring{\mu}^\pm_{\sigma }}{2}\int_{-1}^{1}
\frac{d\tau}{\left( \tau +1\right) ^{2}}\text{ exp}\left[ 
\frac{\varsigma i(\tau-1)}{\tau +1}U^{\beta \gamma }z_{\beta }z_{\gamma }\right] 
\right|_{\rm pole}
\notag \\
&=&- \frac{ \mathring{\mu}^\pm_{\sigma }z_{\alpha }}{4U^{\beta \gamma }z_{\beta }z_{\gamma }} 
\text{ \ \ for \ \ Re}\left( 2\varsigma iU^{\beta \gamma }z_{\beta
}z_{\gamma }\right) >0\text{ .}
\end{eqnarray}
Indeed, taking the exterior derivative of the right-hand side 
one obtains a delta function on the 
holomorphic slice of ${\cal Z}_4$ that cancels the linear source term 
in Eq. \eqref{3.27}.
As for the higher order corrections to $\mathring a_\a$
in the $\nu$-expansion,
they are finite but not analytic at $z^\a=0$, given
by combinations of positive powers and
logarithms of $z^\a$.

As we shall see in Section \ref{Sec analyticity}, the nature of the twistor space 
connection as a distribution on ${\cal Y}_4\times {\cal Z}_4$,
changes drastically once the vacuum gauge function is switched 
on and the connection is given in normal order.

\subsection{Singularities in $L$-gauge from $T_{\vec 0}$}\label{Sec singular id}

We note that if the unity $T_{\vec 0}$ of the star product algebra ${\cal A}_{\Lambda}$ in \eqref{calALambda} is 
represented by the constant symbol on ${\cal Y}_4$, then its contributions to both $V^{(L)}_\alpha$ 
and $\Phi^{(L)}$ that are not real-analytic at the origin of
${\cal Y}_4\times {\cal Z}_4$ for generic points in 
${\cal X}_4$.
More precisely, the singular contributions to $V^{(L)}_\alpha$
are given by $\Pi_\sigma \star a_{\sigma;\vec 0;\alpha}$,
where $a_{\sigma;\vec 0;\alpha}$ is given by 
$\vec\zeta$-independent contribution to \eqref{a-circle soln};
and those to $\Phi^{(L)}$ are given by 
$\nu_{\sigma;\vec 0} \Pi_\sigma\star \kappa_y$.
They are hence singular at $Z^{\underline\alpha}=0$
and $Y^{\underline\alpha}=0$, respectively.
Thus, in order for a star product algebra to give rise
to proper higher spin gravity configurations,
it cannot contain the constant symbol on ${\cal Y}_4$;
in the case of a group algebra this can be achieved by
a truncation to a proper semigroup algebra 
(without the unity), as we shall analyse in
more detail in Sections \ref{Sec Weyl tensor} and \ref{Sec analyticity}.

In the remainder of this section, however, we 
shall proceed with the construction of solution
spaces in the holomorphic gauge without truncating the underlying 
group algebras.

\subsection{Abelian group algebra from Cartan subalgebra of ${\rm sp}(4;\Real)$}

In what follows, we shall give an explicit example of a 
solution space of the type introduced above in the case
when the lattice is two-dimensional, \emph{i.e.} $\vec{\l}=(m,\tilde m)$
with $m,\tilde m\in \mathbb Z$.
The underlying group algebra $\Comp[\mathbb{Z}\times \mathbb Z]$ 
is realized as
\be {\cal A}_{E,J}:= \bigoplus_{\s=\pm} {\cal A}_{E,J;\s}\ ,\qquad
{\cal A}_{E,J;\sigma}:=
\bigoplus_{m,\tilde m\in\mathbb{Z}} \left( T_{m,\tilde{m}}\star \Pi_\s\right)\otimes\Comp
\ ,\ee
in terms of group elements  
\begin{equation}
T_{m,\tilde{m}}:=e_{\star }^{-4m\theta E}\star e_{\star }^{-4\tilde{m}\tilde{
\theta} J}\ , \label{Tdef}
\end{equation}
generated by the anti-de Sitter energy and 
spin operators\footnote{%
Inequivalent exact solution spaces can be obtained
by replacing $E$ and $J$ by Cartan subalgebra
generators in ${\rm sp}(4;\Comp)$, which we leave for future work.} 
\begin{eqnarray}
E &=&\frac{1}{8}E_{\underline{\alpha \beta }}Y^{\underline{\alpha }}\star Y^{%
\underline{\beta }}=\frac{1}{8}E_{\underline{\alpha \beta }}Y^{\underline{%
\alpha }}Y^{\underline{\beta }}\text{ \ ,} \\
J &=&\frac{1}{8}J_{\underline{\alpha \beta }}Y^{\underline{\alpha }%
}\star Y^{\underline{\beta }}=\frac{1}{8}J_{\underline{\alpha \beta }}Y^{%
\underline{\alpha }}Y^{\underline{\beta }}\text{ \ ,}
\end{eqnarray}
respectively, using ${\rm sp}(4;\Real)$ valued matrices obeying%
\footnote{We have suppressed the dummy indices, which are contracted using the north-west to south-east convention. 
}
\be
(E^2)_{\underline{\alpha}}{}^{\underline{\beta }}=(J^2)_{\underline{
\alpha}}{}^{\underline{\beta }}=-\delta _{\underline{\alpha }}{}^{\underline{\beta }}\text{ ,} 
\label{EJProp1} \ee
\be 
(EJ)_{\underline{\alpha }}{}^{\underline{\b}}=(JE)_{\underline{\alpha }}{}^{\underline{\beta }}\text{ ,} \qquad 
(EJ)_{\underline{\alpha\beta }}+(JE)_{\underline{\beta\alpha }}=0\text{ ,}  \qquad
(EJ)_{\underline{\alpha }}{}^{\underline{\a}}=0\ , 
\label{EJProp2} 
\ee
from which it follows that
\be \text{det}\left( \delta _{\underline{\alpha }}{}^{\underline{\beta }}+aE_{%
\underline{\alpha }}{}^{\underline{\gamma }}J_{\underline{\gamma }}{}^{%
\underline{\beta }}\right) =\left( 1-a^{2}\right) ^{2}\text{ .}
\label{EJProp4}\ee
As for the parameters, we take
\begin{equation}
\theta\in \mathbb{R}\cup i\mathbb{R}\ ,\qquad 
i\mathbb{Z} \theta \cap \left( \frac{\pi }{2}+\mathbb{Z\pi }\right)
=\emptyset \text{ ,}
\end{equation}
\emph{idem} $\tilde{\theta}$.
The basis elements obey \eqref{basis}, viz.
\begin{equation}
T_{m,\tilde{m}}\star T_{n,\tilde{n}}=T_{m+n,\tilde{m}+\tilde{n}}\text{ ,}\qquad 
\left[ T_{m,\tilde{m}},\Pi _{\sigma }\right] _{\star }=0\text{ ,}
\qquad \pi(T_{m,\tilde{m}})=T_{-m,\tilde{m}}\ .
\end{equation}

To compute the symbol of $T_{m,\tilde m}$ in Weyl order,
we first use (\ref{star and ordinary exponent}) with $N=4$ 
to compute
\begin{equation}
e_{\star }^{-4m\theta E}=\mathbf{S}^{2}\,
 e^{-4\mathbf{T} E} \text{ \ , \ \ }e_{\star }^{-4\tilde{m}\tilde{
\theta}{J}}=
\widetilde{\mathbf{S}}^{2} \,e^{-4\widetilde{\mathbf{T}} J}\text{ ,}
\label{star and ordinary exponent E J}
\end{equation}
where
\begin{equation}
\mathbf{S}:= \text{sech}\left( m\theta \right) \text{ , \quad }\mathbf{T}
:= \text{tanh}\left( m\theta \right) \text{ , \quad }\widetilde{\mathbf{S}}
:= \text{sech}\left( \tilde{m}\tilde{\theta}\right) \text{ , \quad }
\widetilde{\mathbf{T}}:= \text{tanh}\left( \tilde{m}\tilde{\theta}
\right) \text{ . \ }
\end{equation}
In what follows, we make the convention that all boldfaced quantities depend on 
$m\theta$ and $\tilde{m}\tilde{\theta}$.
The symbol of $T_{m,\tilde{m}}$ is thus given by 
\begin{eqnarray}
T_{m,\tilde{m}} &=&\left[ \mathbf{S}^{2}e^{-4\mathbf{T}E}\right] \star \left[
\widetilde{\mathbf{S}}^{2}e^{-4\widetilde{\mathbf{T}}J}\right]
\notag \\
&=&\left( \mathbf{S}\widetilde{\mathbf{S}}\right) ^{2}\int \frac{d^{4}Ud^{4}V%
}{\left( 2\pi \right) ^{4}}\text{ exp}\left\{ i\left( V^{\underline{\alpha }%
}-Y^{\underline{\alpha }}\right) \left( U_{\underline{\alpha }}-Y_{%
\underline{\alpha }}\right) \right\}  \notag \\
&&\ \ \ \ \ \ \ \ \ \ \ \ \ \ \ \ \ \ \ \ \ \ \ \ \ \ \ \ \ \ \text{exp}%
\left\{ -\frac{1}{2}\left[ \mathbf{T}E_{\underline{\alpha \beta }}U^{%
\underline{\alpha }}U^{\underline{\beta }}+\widetilde{\mathbf{T}}J_{%
\underline{\alpha \beta }}V^{\underline{\alpha }}V^{\underline{\beta }}%
\right] \right\} \text{ .}
\end{eqnarray}
By performing the Gaussian integrals, we obtain
\begin{equation}
T_{m,\tilde{m}}=\mathbf{A}\text{exp}\left\{ -\frac{1}{2} \mathbf{K}_{\underline{\alpha \beta }} Y^{\underline{\alpha }}Y^{\underline{\beta }}\right\} \text{ ,}\qquad
\mathbf{K}_{\underline{\alpha \beta }}:=\mathbf{B}E_{\underline{
\alpha \beta }}+\mathbf{C}J_{\underline{\alpha \beta }}\text{ ,}
\label{Tmm EJ ABC}
\end{equation}
where
\begin{equation}
\mathbf{A}:= \frac{\left( \mathbf{S}\widetilde{\mathbf{S}}\right) ^{2}}{%
1-\left( \mathbf{T}\widetilde{\mathbf{T}}\right) ^{2}}\text{ , \quad }\mathbf{B}%
:=\frac{\mathbf{T}\left( 1-\widetilde{\mathbf{T}}^{2}\right) }{1-\left(
\mathbf{T}\widetilde{\mathbf{T}}\right) ^{2}}\text{ , \quad }\mathbf{C}:=
\frac{\widetilde{\mathbf{T}}\left( 1-\mathbf{T}^{2}\right) }{1-\left(
\mathbf{T}\widetilde{\mathbf{T}}\right) ^{2}}\text{ .}  \label{ABC notation}
\end{equation}
We also need the symbol of $T_{m,\tilde{m}}\star \kappa_{y}\bar{\kappa}_{\bar{y}}$, 
which is given by
\begin{eqnarray}
T_{m,\tilde{m}}\star \kappa _{y}\bar{\kappa}_{\bar{y}}  \notag 
&=&\left( 2\pi \right) ^{2}\mathbf{A}\text{exp}\left\{ -\frac{1}{2}\mathbf{K}%
_{\underline{\alpha \beta }}Y^{\underline{\alpha }}Y^{\underline{\beta }%
}\right\} \star \delta ^{4}\left( Y\right)  \notag \\
&=&\mathbf{A}\int \frac{d^{4}Ud^{4}V}{\left( 2\pi \right) ^{2}}e^{iV^{%
\underline{\alpha }}U_{\underline{\alpha }}}e^{-\frac{1}{2}\mathbf{K}_{%
\underline{\alpha \beta }}\left( Y^{\underline{\alpha }}+U^{\underline{%
\alpha }}\right) \left( Y^{\underline{\beta }}+U^{\underline{\beta }}\right)
}\delta ^{4}\left( Y+V\right)  \notag \\
&=&\mathbf{A}\int \frac{d^{4}Ud^{4}V}{\left( 2\pi \right) ^{2}}e^{-iY^{%
\underline{\alpha }}U_{\underline{\alpha }}}e^{-\frac{1}{2}\mathbf{K}_{%
\underline{\alpha \beta }}U^{\underline{\alpha }}U^{\underline{\beta }}}
\notag \\
&=&\frac{\mathbf{A}}{\sqrt{\text{det}\left( \mathbf{K}\right) }}\text{exp}%
\left\{ -\frac{1}{2}\left( \mathbf{K}^{-1}\right) ^{\underline{\alpha \beta }%
}Y_{\underline{\alpha }}Y_{\underline{\beta }}\right\} \text{ ,}
\label{Tkappa}
\end{eqnarray}
where 
\begin{equation}
\mathbf{K}_{\underline{\alpha \beta }}\left( \mathbf{K}^{-1}\right) ^{%
\underline{\beta \gamma }}:=\delta _{\underline\alpha }{}^{\underline\gamma }\text{ .}
\end{equation}

\subsection{New exact biaxially symmetric solutions in holomorphic gauge}\label{Sec biaxial sym}

The above construction of ${\cal A}_{E,J}$ thus allows us to
solve equations (\ref{X-indep-VJ})--(\ref{X-indep-V}) 
using the Ansatz \eqref{Phiprime}--\eqref{Vprime}.
In order to keep matters simple, we shall assume 
that $\check \nu = \check \a = 0$, and work
with the following reduced version:
\begin{eqnarray}
\Phi ^{\prime } &=&\sum_{\sigma ;m,\tilde{m}}\nu _{\sigma ;m,\tilde{m}}T_{m,%
\tilde{m}}\star \Pi _{\sigma }\star \kappa _{y}\text{ ,}  \label{ansatzF_YZ}
\\
V_{\alpha }^{\prime } &=&\sum_{\sigma ;m,\tilde{m}}T_{m,\tilde{m}}\star \Pi
_{\sigma }\star \left( a_{\sigma ;m,\tilde{m}}(z)\right) _{\alpha }\text{ ,}
\label{ansatzV_YZ}\\
\bar{V}_{\dot{\alpha}}^{\prime }&=&(V_{\alpha }^{\prime })^\dag=\sum_{\sigma ;m,\tilde{m}}T_{m,\tilde{m}%
}^{\dag }\star \Pi _{\sigma }\star \left( \bar{a}_{\sigma ;m,\tilde{m}
}(\bar z)\right) _{\dot{\alpha}}\text{ , }
\label{ansatzVbar_YZ}
\end{eqnarray}
where thus $\nu _{\sigma;m,\tilde{m}}\in \mathbb{C}$ and we recall that 
the twistor space connection is (anti-)holomorphic, as indicated above.
From 
\begin{equation}
\begin{tabular}{c|cc}
& $\theta \in \mathbb{R}$ & $\theta \in i\mathbb{R}$ \\ \hline
$\tilde{\theta}\in \mathbb{R}$ & $T_{m,\tilde{m}}^{\dag }=T_{m,\tilde{m}}$ &
$T_{m,\tilde{m}}^{\dag }=T_{-m,\tilde{m}}$ \\
$\tilde{\theta}\in i\mathbb{R}$ & $T_{m,\tilde{m}}^{\dag }=T_{m,-\tilde{m}}$
& $T_{m,\tilde{m}}^{\dag }=T_{-m,-\tilde{m}}$
\end{tabular}
\end{equation}
it follows that the reality condition $(\Phi')^\dag=\pi(\Phi')$ implies that 
\begin{equation}
\begin{tabular}{c|cc}
& $\theta \in \mathbb{R}$ & $\theta \in i\mathbb{R}$ \\ \hline
$\tilde{\theta}\in \mathbb{R}$ & $\nu _{\sigma ;m,\tilde{m}}^{\ast }=\sigma
\nu _{\sigma ;m,\tilde{m}}$ & $\nu _{\sigma ;m,\tilde{m}}^{\ast }=\sigma \nu
_{\sigma ;-m,\tilde{m}}$ \\
$\tilde{\theta}\in i\mathbb{R}$ & $\nu _{\sigma ;m,\tilde{m}}^{\ast }=\sigma
\nu _{\sigma ;m,-\tilde{m}}$ & $\nu _{\sigma ;m,\tilde{m}}^{\ast }=\sigma
\nu _{\sigma ;-m,-\tilde{m}}$
\end{tabular}\text{\ \ .}
\label{realcond-nu-sigma}
\end{equation}
We note that the Ansatz \eqref{ansatzF_YZ}--\eqref{ansatzVbar_YZ} identically obeys (\ref{X-indep-PhiV}) and (\ref{X-indep-V}) 
since \be \left[ \left( a_{\sigma ;m,\tilde{m}}\right) _{\alpha },\left( \bar{a}
_{\sigma ;m,\tilde{m}}\right) _{\dot{\alpha}}\right] _{\star }=0 \ ,\ee
while (\ref{X-indep-VJ}) reduces to 
\begin{equation}
\partial _{\lbrack \alpha }\left( a_{\sigma ;m,\tilde{m}}\right) _{\beta
]}+\sum_{n,\tilde{n}}\left( a_{\sigma ;n,\tilde{n}}\right) _{[\alpha }\star
\left( a_{\sigma ;m-n,\tilde{m}-\tilde{n}}\right) _{\beta ]}+\frac{i}{4}%
\varepsilon _{\alpha \beta }\mu _{\sigma ;m,\tilde{m}}\kappa _{z}=0\text{ \
and h.c.\ ,}  \label{z-dependent eq}
\end{equation}
where $\mu _{\sigma ;m,\tilde{m}}$ are defined as in \eqref{defmu}.
Finally, multiplying (\ref{z-dependent eq}) with $\zeta ^{m}\tilde{\zeta} ^{\tilde{m}}$,
where $\zeta,\tilde{\zeta}\in \mathbb{C}$, and summing over $m$ 
and $\tilde{m}$, yields the equivalent equation
\begin{equation}
\partial _{\lbrack \alpha }\left( \mathring{a}_{\sigma }\right) _{\beta
]}+\left( \mathring{a}_{\sigma }\right) _{[\alpha }\star \left( \mathring{a}
_{\sigma }\right) _{\beta ]}+\frac{i}{4}\varepsilon _{\alpha \beta }
\mathring{\mu}_{\sigma }\kappa _{z}=0\text{ \ and h.c.\ ,}
\label{z-dependent eq simp}
\end{equation}
where the generating functions 
\begin{equation}
\left( \mathring{a}_{\sigma }\right) _{\alpha }\left( \zeta ,\tilde{\zeta}
\right) := \sum_{m,\tilde{m}}\left( {a}_{\sigma ,m,\tilde{m}
}\right) _{\alpha }\zeta ^{m}\tilde\zeta ^{\tilde{m}}\text{\ \ , \ \ }\mathring{\mu
}_{\sigma }\left( \zeta ,\tilde{\zeta}\right) := \sum_{m,\tilde{m}}\mu
_{\sigma ;m,\tilde{m}}\zeta ^{m}\tilde\zeta ^{\tilde{m}}\text{ ,}
\end{equation}
for which we shall use the particular solutions in \eqref{a-circle soln}
with $\check \nu = \check a = 0$.

By definition, the symmetries of the solution are generated by 
generalized Killing gauge parameters $\epsilon^{(G)}$ leaving 
$(\Phi^{(G)},U_\mu^{(G)},V_\alpha^{(G)})$ invariant.
Locally, the space of such parameters is given by
\be \epsilon^{(G)}= G^{-1}\star \epsilon' \star G\ ,\qquad
\epsilon'=\epsilon'(E,J)\ ,\ee
where the parameters are arbitrary star polynomials in $E$
and $J$;
and globally, a Killing parameter belongs to an adjoint section 
obeying suitable boundary conditions, and we shall assume
that $\epsilon^{(G)}$ is
real-analytic on ${\cal Y}_4\times {\cal Z}_4$ and falls
off at infinity of ${\cal X}_4$, such that they leave the
background spacetime invariant.
This implies that the solutions have time-translational 
and rotational symmetries generated by $E$ and $J$, respectively.
Furthermore, if the Ansatz is expanded over only $T_{m,0}$ or
$T_{0,\tilde m}$, respectively, then the symmetry is further 
enhanced to the enveloping algebras of ${\rm so}(2)_E \oplus {\rm so}(3)$
or ${\rm so}(1,2)\oplus {\rm so}(2)_J$, where ${\rm so}(3)$ is the 
subalgebras of ${\rm sp}(4;\Real)$ commuting to $E$ \emph{idem} 
${\rm so}(1,2)$ and $J$.
Acting on the solutions with the full higher spin algebra leads to
an orbit that forms a higher spin representation space.  
The trace operation ${\rm Tr}'$ equips this space with an indefinite
sesqui-linear form, as we shall comment on below in the context of 
higher spin invariant functionals.

\section{Weyl zero-form and Weyl tensors}\label{Sec Weyl tensor}

In this section we compute the Weyl zero-form, Weyl tensors and higher spin invariants formed out of them.

\subsection{The Weyl zero-form in $L$-gauge}

The Weyl tensors in the $L$-gauge are contained in the 
zero-form master field. From (\ref{AnsatzPhi}) and (\ref{ansatzF_YZ}) it follows
that 
\begin{eqnarray}
\Phi^{(L)}&=& \frac{1}{2}\sum_{\sigma ,m,\tilde{m}}\nu _{\sigma ,m,\tilde{m}%
}L^{-1}\star T_{m,\tilde{m}}\star \left( \kappa _{y}+\sigma \bar{\kappa}_{%
\bar{y}}\right) \star \pi \left( L\right) \text{ ,}\notag\\
 &=&\frac{1}{2}\sum_{\sigma ,m,\tilde{m}}\nu _{\sigma ,m,\tilde{m}%
}\left( L^{-1}\star T_{m,\tilde{m}}\star L\right) \star \left( \kappa
_{y}+\sigma \bar{\kappa}_{\bar{y}}\right)  \notag \\
&=&\sum_{m,\tilde{m}}\left( \nu _{1,m,\tilde{m}}T_{m,\tilde{m}}^L\star \kappa _{y}+\nu _{2,m,\tilde{m}}T_{m,\tilde{m}}^L\star \bar{\kappa}_{\bar{y}}\right) \text{ ,}  \label{Phi two terms}
\end{eqnarray}
where
\be T_{m,\tilde{m}}^L:=L^{-1}\star T_{m,\tilde{m}}\star L\ ,\ee
and the parameters 
\begin{equation}
\nu _{1,m,\tilde{m}} := \frac{1}{2}\left( \nu _{+,m,\tilde{m}}+\nu _{-,m,%
\tilde{m}}\right) \text{ \ , \ }\nu _{2,m,\tilde{m}} := \frac{1}{2}\left(
\nu _{+,m,\tilde{m}}-\nu _{-,m,\tilde{m}}\right) \text{ ,}\label{nurelation}
\end{equation}
obey the reality conditions 
\begin{equation}
\begin{tabular}{c|cc}
& $\theta \in \mathbb{R}$ & $\theta \in i\mathbb{R}$ \\ \hline
$\tilde{\theta}\in \mathbb{R}$ & $\nu _{1,m,\tilde{m}}^{\ast }=\nu _{2,m,%
\tilde{m}}$ & $\nu _{1,m,\tilde{m}}^{\ast }=\nu _{2,-m,\tilde{m}}$ \\
$\tilde{\theta}\in i\mathbb{R}$ & $\nu _{1,m,\tilde{m}}^{\ast }=\nu _{2,m,-%
\tilde{m}}$ & $\nu _{1,m,\tilde{m}}^{\ast }=\nu _{2,-m,-\tilde{m}}$%
\end{tabular}
\end{equation}
To compute $ T_{m,\tilde{m}}^L$ we use the lemma
\begin{equation}
L^{-1}\star f\left( Y_{\underline{\alpha }}\right) \star L=f\left( L_{%
\underline{\alpha }}{}^{\underline{\beta }}Y_{\underline{\beta }}\right)
\text{ ,}
\end{equation}
where $L_{\underline{\alpha }}{}^{\underline{\beta }}$ is a matrix that
depends on the spacetime coordinates (see Appendix \ref{Sec coordinates} for
an explicit expression).
It follows from (\ref{Tmm EJ ABC}) that 
\begin{equation}
T_{m,\tilde{m}}^L=\mathbf{A}\text{exp}\left\{ \left( -\frac{%
1}{2}\right) \mathbf{K}_{\underline{\alpha \beta }}^{L} Y^{\underline{\alpha }}Y^{\underline{%
\beta }}\right\} \text{ ,}  \label{LTmL EJ ABC}
\end{equation}
where 
\begin{equation}
 \mathbf{K}_{\underline{\alpha \beta }}^{L}:= 
 \mathbf{B}E_{\underline{\alpha \beta }}^{L}+\mathbf{C}J_{
\underline{\alpha \beta }}^{L}\ ,\qquad
E^{L}_{\underline{\alpha \beta }}:= E_{\underline{\gamma
\delta }}L^{\underline{\gamma }}{}_{\underline{\alpha }}L^{\underline{\delta
}}{}_{\underline{\beta }}\text{ , \qquad}J^{L}_{\underline{\alpha
\beta }}:=J_{\underline{\gamma \delta }}L^{\underline{\gamma }}{}_{
\underline{\alpha }}L^{\underline{\delta }}{}_{\underline{\beta }}\text{ .}
\end{equation}
Under $Y^{\underline{\alpha }}=\left\{ y^{\alpha },
\bar{y}^{\dot{\alpha}}\right\} $, the above
matrices decompose into
\begin{equation}
E_{\underline{\alpha \beta }}^{L}=:\left(
\begin{array}{cc}
(\kappa_E^{L})_{\alpha \beta } & (v_E^{L})_{\alpha \dot{\beta}} \\
(\bar{v}_E^{L})_{\dot{\alpha}\beta } & (\bar{\kappa}_E^{L})_{\dot{\alpha}\dot{\beta}}
\end{array}
\right) \text{ ,}
\end{equation}
\emph{idem} $J_{\underline{\alpha \beta }}^{L}$, whose components 
obey%
\footnote{Eqs.\ \eqref{v-prop}--\eqref{kk-kvv} hold, if we label all components with either ``$E$'' or ``$J$'' (not a mixture of both). }
\begin{equation}
v_{\alpha \dot{\beta}}^{L}=\bar{v}_{\dot{\beta}\alpha }^{L}\text{ , \qquad }
\left( v^{L}\right) _{\alpha \dot{\beta}}\left(\bar v^{L}\right) ^{\dot{\beta}
\gamma }=\left( v^{L}\right) ^{2}\delta _{\alpha }{}^{\gamma }\text{ , \qquad }
\left( \bar{v}^{L}\right) _{\dot{\alpha}\beta }\left( v^{L}\right)
^{\beta \dot{\gamma}}=\left( v^{L}\right) ^{2}\delta _{\dot \alpha }{}^{\dot \gamma }
\text{ ,}
\label{v-prop}
\end{equation}
where $\left( v^{L}\right)^{2}:= \frac{1}{2}\left( v^{L}\right) _{\alpha \dot{\beta}%
}\left( v^{L}\right) ^{\alpha \dot{\beta}}$,
and
\begin{equation}
\left(\kappa^L\right)^2:= \frac{1}{2}\left(\kappa^L\right)_{\alpha \beta }\left(\kappa^L\right)^{\alpha
	\beta }=\text{det}\left(\kappa^L\right)
\ \ \ \ , \ \ \ \ \ \ \ 
\left(\kappa^L\right)_{\alpha\beta }\left(\kappa^L\right)^{\beta \gamma }=\left(\kappa^L\right)^{2}\delta _{\alpha }{}^{\gamma } \ \ ,
\end{equation}
\emph{idem} $\bar\kappa^L$, which are derived from general properties of any 2$\times$2 symmetric matrix.
Furthermore, from \eqref{EJProp1} it follows that
\begin{equation}
\left( \kappa ^{L}\right) _{\alpha \beta }\left( v^{L}\right) ^{\beta \dot{%
\gamma}}+\left( v^{L}\right) _{\alpha \dot{\beta}}\left( \bar{\kappa}%
^{L}\right) ^{\dot{\beta}\dot{\gamma}}=0\text{ , \qquad }\left( \kappa
^{L}\right) ^{2}+\left( v^{L}\right) ^{2}=\left( \bar{\kappa}^{L}\right)
^{2}+\left( v^{L}\right) ^{2}=1\text{ ,}
\end{equation}
which in its turn implies 
\begin{equation}
\left( \bar{\kappa}^{L}\right) ^{2}\left( \kappa ^{L}\right) _{\alpha \beta
}-\left( \bar{\kappa}^{L}\right) ^{\dot{\alpha}\dot{\beta}}\left(
v^{L}\right) _{\alpha \dot{\alpha}}\left( v^{L}\right) _{\beta \dot{\beta}%
}=\left( \kappa ^{L}\right) _{\alpha \beta }\text{ ,}  \label{kk-kvv}
\end{equation}
which will be useful later when we determine the Petrov type.

Returning to (\ref{LTmL EJ ABC}), we thus have
\begin{equation}
\mathbf{K}_{\underline{\alpha \beta }}^{L} = \left(
\begin{array}{cc}
\mathbf{F}_{\alpha \beta } & \mathbf{G}_{\alpha \dot{\beta}} \\
\mathbf{G}_{\dot{\alpha}\beta } & \mathbf{H}_{\dot{\alpha}\dot{\beta}}%
\end{array}%
\right) :=\left(
\begin{array}{cc}
\mathbf{B}\left( \kappa _{E}^{L}\right) _{\alpha \beta }+\mathbf{C}\left(
\kappa _{J}^{L}\right) _{\alpha \beta } & \mathbf{B}\left( v_{E}^{L}\right)
_{\alpha \dot{\beta}}+\mathbf{C}\left( v_{J}^{L}\right) _{\alpha \dot{\beta}}
\\
\mathbf{B}\left( \bar{v}_{E}^{L}\right) _{\dot{\alpha}\beta }+\mathbf{C}
\left( \bar{v}_{J}^{L}\right) _{\dot{\alpha}\beta } & \mathbf{B}\left( \bar{
\kappa}_{E}^{L}\right) _{\dot{\alpha}\dot{\beta}}+\mathbf{C}\left( \bar{
\kappa}_{J}^{L}\right) _{\dot{\alpha}\dot{\beta}}
\end{array}
\right) \text{ ,}
\end{equation}
where $\mathbf{G}_{\dot{\alpha}\beta }=\mathbf{G}_{\beta\dot{\alpha}}$,
and correspondingly
\begin{equation}
T_{m,\tilde{m}}^L=\mathbf{A}\text{exp}\left\{ \left( -\frac{%
1}{2}\right) \left[ y^{\alpha }\mathbf{F}_{\alpha \beta }y^{\beta }+\bar{y}^{%
\dot{\alpha}}\mathbf{H}_{\dot{\alpha}\dot{\beta}}\bar{y}^{\dot{\beta}%
}+2y^{\alpha }\mathbf{G}_{\alpha \dot{\beta}}\bar{y}^{\dot{\beta}}\right]
\right\} \text{ .}
\end{equation}
Finally, for $(m,\tilde m)\neq (0,0)$, by performing Gaussian integrals 
we obtain%
\footnote{We note the useful
	relations 
	$\mathbf{F}^{2}:= \frac{1}{2}\mathbf{F}_{\alpha \beta }(\mathbf{F})^{\alpha
		\beta }=\text{det}\left( \mathbf{F}\right)$ and $
	\mathbf{F}_{\alpha
		\beta }\mathbf{F}^{\beta \gamma }=\mathbf{F}^{2}\delta _{\alpha }{}^{\gamma }$, \emph{idem} $\mathbf{H}$.}
\begin{eqnarray}
&&T_{m,\tilde{m}}^L\star \kappa _{y}  \notag \\
&=&\!\!\!\!\frac{\mathbf{A}}{\sqrt{\mathbf{F}^{2}}}\text{exp}\left\{ \frac{1}{2%
\mathbf{F}^{2}}\left[ \left( \mathbf{F}^{\alpha \beta }\mathbf{G}_{\alpha
\dot{\alpha}}\mathbf{G}_{\beta \dot{\beta}}-\mathbf{F}^{2}\mathbf{H}_{\dot{%
\alpha}\dot{\beta}}\right) \bar{y}^{\dot{\alpha}}\bar{y}^{\dot{\beta}}-%
\mathbf{F}_{\alpha \beta }y^{\alpha }y^{\beta }+2i\mathbf{F}_{\alpha
}{}^{\beta }\mathbf{G}_{\beta \dot{\beta}}y^{\alpha }\bar{y}^{\dot{\beta}}%
\right] \right\} \text{ ,}  \notag \\
&&
\end{eqnarray}
and
\begin{eqnarray}
&&T_{m,\tilde{m}}^L\star \bar{\kappa}_{\bar{y}}  \notag \\
&=&\!\!\!\!\frac{\mathbf{A}}{\sqrt{\mathbf{H}^{2}}}\text{exp}\left\{ \frac{1}{2%
\mathbf{H}^{2}}\left[ \left( \mathbf{H}^{\alpha \beta }\mathbf{G}_{\alpha
\dot{\alpha}}\mathbf{G}_{\beta \dot{\beta}}-\mathbf{H}^{2}\mathbf{F}_{\dot{%
\alpha}\dot{\beta}}\right) y^{\alpha }y^{\beta }-\mathbf{H}_{\dot{\alpha}%
\dot{\beta}}\bar{y}^{\dot{\alpha}}\bar{y}^{\dot{\beta}}+2i\mathbf{H}_{\dot{%
\alpha}}{}^{\dot{\beta}}\mathbf{G}_{\beta \dot{\beta}}\bar{y}^{\dot{\alpha}%
}y^{\beta }\right] \right\} \text{ ,}  \notag \\
&&
\end{eqnarray}
while $T_{0,\tilde{0}}^L\star \kappa_{y}=  \kappa_{y}$ and  $T_{0,\tilde{0}}^L\star \bar\kappa_{\bar y}=  \bar \kappa_{\bar y}$.

Substituting the above formulae into (\ref{Phi two terms}), we obtain
\begin{align}
&\Phi^{(L)} \notag \\
=& \ \nu_{1,0,0}\kappa_{y}\ +\ \nu_{2,0,0} \bar \kappa_{\bar y}\ +
\notag\\
&\sum_{(m,\tilde{m})\neq (0,0)}\mathbf{A}\left( \frac{\nu _{1,m,\tilde{m}}}{\sqrt{%
\mathbf{F}^{2}}}\text{exp}\left\{ \frac{1}{2\mathbf{F}^{2}}\left[ \left(
\mathbf{F}^{\alpha \beta }\mathbf{G}_{\alpha \dot{\alpha}}\mathbf{G}_{\beta
\dot{\beta}}-\mathbf{F}^{2}\mathbf{H}_{\dot{\alpha}\dot{\beta}}\right) \bar{y%
}^{\dot{\alpha}}\bar{y}^{\dot{\beta}}-\mathbf{F}_{\alpha \beta }y^{\alpha
}y^{\beta }+2i\mathbf{F}_{\alpha }{}^{\beta }\mathbf{G}_{\beta \dot{\beta}%
}y^{\alpha }\bar{y}^{\dot{\beta}}\right] \right\} \right.  \notag \\
& \ \ \ \left. +\frac{\nu _{2,m,\tilde{m}}}{\sqrt{\mathbf{H}^{2}}}%
\text{exp}\left\{ \frac{1}{2\mathbf{H}^{2}}\left[ \left( \mathbf{H}^{\dot{%
\alpha}\dot{\beta}}\mathbf{G}_{\alpha \dot{\alpha}}\mathbf{G}_{\beta \dot{%
\beta}}-\mathbf{H}^{2}\mathbf{F}_{\alpha \beta }\right) y^{\alpha }y^{\beta
}-\mathbf{H}_{\dot{\alpha}\dot{\beta}}\bar{y}^{\dot{\alpha}}\bar{y}^{\dot{%
\beta}}+2i\mathbf{H}_{\dot{\alpha}}{}^{\dot{\beta}}\mathbf{G}_{\beta \dot{%
\beta}}\bar{y}^{\dot{\alpha}}y^{\beta }\right] \right\} \right) \text{ .}
\notag \\
&
\end{align}
The expression $\mathbf{H}^{2}\mathbf{F}_{\alpha\beta}-
\mathbf{H}^{\dot{\alpha}\dot{\beta}}\mathbf{G}_{\alpha\dot{\alpha}}
\mathbf{G}_{\beta\dot{\beta}}$ can be factorized as
\begin{equation}
\mathbf{H}^{2}\mathbf{F}_{\alpha\beta}-
\mathbf{H}^{\dot{\alpha}\dot{\beta}}\mathbf{G}_{\alpha\dot{\alpha}}
\mathbf{G}_{\beta\dot{\beta}}\ =\ \left(\mathbf{B}^2-\mathbf{C}^2\right)\mathbf{\breve F}_{\a\b} \ ,
\end{equation}
where $\mathbf{\breve F}_{\a\b}$ satisfies
\be
\mathbf{\breve F}^2\ =\ \mathbf{H}^2 \ .
\ee
Then, assuming that 
\be \nu_{1,0,0}=0=\nu_{2,0,0}\ ,\ee
the resulting generalized spin-$s$ Weyl tensor in the $L$-gauge reads
\begin{eqnarray}
&&C_{\alpha _{1}\cdots \alpha _{2s}}  \notag \\
&:=&\left[ \frac{\partial }{\partial y^{\alpha _{1}}}\cdots \frac{\partial }{%
\partial y^{\alpha _{2s}}}\Phi^{(L)} \right] _{Y=0}  \notag \\
&=&\frac{\left( 2s\right) !}{s!}\sum_{(m,\tilde{m})\neq (0,0)}\mathbf{A}\left\{ \frac{%
\nu _{1,m,\tilde{m}}}{\sqrt{\mathbf{F}^{2}}}\left( \frac{-1}{2\mathbf{F}^{2}}%
\right) ^{s}\mathbf{F}_{(\alpha _{1}\alpha _{2}}\cdots \mathbf{F}_{\alpha
_{2s-1}\alpha _{2s})}\right.  \notag \\
&&\ \ \ \ \ \ \ \ \ \ \ \ \ \  
\left.+\left(\mathbf{B}^2-\mathbf{C}^2\right)^{s} \frac{\nu _{2,m,\tilde{m}}}{\sqrt{%
\mathbf{\mathbf{\breve{F}}}^{2}}}\left( \frac{-1}{2\mathbf{\breve{F}}^{2}}\right) ^{s}
\mathbf{\breve F}_{(\alpha _{1}\alpha _{2}}\cdots \mathbf{\breve F}_{\alpha
_{2s-1}\alpha _{2s})}
\right\} \text{ ,} 
\label{Weyl tensor}
\end{eqnarray}
where there are two separate generalized Petrov type-D tensors summed for each $(m, \tilde{m})$ for positive $s$.%
\footnote{A generalized spin-$s$ Petrov type-D tensor is defined as a symmetric rank-$2s$ tensor with spinor indices that can be decomposed into the products of two spinors, each of which has the power $s$  \cite{Iazeolla:2011cb}.}

\subsection{Petrov types of the Weyl tensors}
In what follows, we analyze in a few special cases whether the Weyl tensor as the sum \eqref{Weyl tensor} is of Petrov type D.

\paragraph{The case $\protect\theta\tilde{%
\protect\theta}=0$.}

If $\theta\neq 0$ and $\tilde{\theta}=0$, then $\widetilde{\mathbf{S}}=1$, $%
\widetilde{\mathbf{T}}=0$ and 
\begin{equation}
\mathbf{A}=\mathbf{S}^{2}\text{ , \ }\mathbf{B}=\mathbf{T}\text{ , \ }%
\mathbf{C}=0\text{ .}
\end{equation}
It follows that
\begin{equation}
\left(
\begin{array}{cc}
\mathbf{F}_{\alpha \beta } & \mathbf{G}_{\alpha \dot{\beta}} \\
\mathbf{G}_{\dot{\alpha}\beta } & \mathbf{H}_{\dot{\alpha}\dot{\beta}}%
\end{array}%
\right) =\mathbf{T}\left(
\begin{array}{cc}
\left( \kappa _{E}^{L}\right) _{\alpha \beta } & \left( v_{E}^{L}\right)
_{\alpha \dot{\beta}} \\
\left( \bar{v}_{E}^{L}\right) _{\dot{\alpha}\beta } & \left( \bar{\kappa}%
_{E}^{L}\right) _{\dot{\alpha}\dot{\beta}}%
\end{array}%
\right) \text{ .}
\end{equation}
Furthermore, using (\ref{kk-kvv}) we obtain
\begin{equation}
\mathbf{\breve F}_{\alpha\beta} =
\mathbf{T} \left( \kappa
_{E}^{L}\right) _{\alpha\beta} \text{ .}
\end{equation}
The resulting spin-$s$ Weyl tensor reads%
\footnote{One can show that $\left(\kappa_E^{L}\right)^2=-\lambda^2 r^2$.
	The analytical continuation involves the choice of sign in front of the square roots.
	These must be correlated to analogous choices in the expression for
	the twistor space connection.  }
\begin{eqnarray*}
&&C_{\alpha _{1}\cdots \alpha _{2s}}|_{\tilde{\theta}=0} \\
&=&\frac{\left( 2s\right) !}{s!}\sum_{m\neq0}\frac{\mathbf{S}^{2}}{\sqrt{\mathbf{T%
}^{2}\left( \kappa _{E}^{L}\right) ^{2}}}\frac{1}{\left[ -2\left( \kappa
_{E}^{L}\right) ^{2}\right] ^{s}}\left( \nu _{1,m}\mathbf{T}^{-s}+\nu _{2,m}%
\mathbf{T}^{s}\right) \left( \kappa _{E}^{L}\right) _{(\alpha _{1}\alpha
_{2}}\cdots \left( \kappa _{E}^{L}\right) _{\alpha _{2s-1}\alpha _{2s})}%
\text{ ,}
\end{eqnarray*}%
\begin{equation}
\end{equation}%
where $\nu _{1,m}=\sum_{\tilde{m}}\nu _{1,m,\tilde{m}}$ and $\nu
_{2,m}=\sum_{\tilde{m}}\nu _{2,m,\tilde{m}}$, which we assume to be
finite and vanishing for $m=0$.

If instead $\theta =0$ and $\tilde{\theta}\neq 0$, then $\mathbf{S}=1$, $\mathbf{T}=0$, and 
\begin{equation}
\mathbf{A}=\widetilde{\mathbf{S}}^{2}\text{ , \ }\mathbf{B}=0\text{ , \ }%
\mathbf{C}=\widetilde{\mathbf{T}}\text{ .}
\end{equation}
Then we have
\begin{equation}
\left(
\begin{array}{cc}
\mathbf{F}_{\alpha \beta } & \mathbf{G}_{\alpha \dot{\beta}} \\
\mathbf{G}_{\dot{\alpha}\beta }^{T} & \mathbf{H}_{\dot{\alpha}\dot{\beta}}%
\end{array}%
\right) =\widetilde{\mathbf{T}}\left(
\begin{array}{cc}
\left( \kappa _{J}^{L}\right) _{\alpha \beta } & \left( v_{J}^{L}\right)
_{\alpha \dot{\beta}} \\
\left( \bar{v}_{J}^{L}\right) _{\dot{\alpha}\beta } & \left( \bar{\kappa}%
_{J}^{L}\right) _{\dot{\alpha}\dot{\beta}}%
\end{array}%
\right) \text{ ,}
\end{equation}
and hence using (\ref{kk-kvv}) it follows that
\begin{equation}
\mathbf{\breve F}_{\alpha\beta} =
-\widetilde{\mathbf{T}} \left( \kappa
_{J}^{L}\right) _{\alpha\beta} \text{ ,}
\end{equation}
and the spin-$s$ Weyl tensor becomes%
\begin{eqnarray*}
&&C_{\alpha _{1}\cdots \alpha _{2s}}|_{\theta =0} \\
&=&\frac{\left( 2s\right) !}{s!}\sum_{\tilde{m}\neq 0}\frac{\widetilde{\mathbf{S}}%
^{2}}{\sqrt{\widetilde{\mathbf{T}}^{2}\left( \kappa _{J}^{L}\right) ^{2}}}%
\frac{1}{\left[ -2\left( \kappa _{J}^{L}\right) ^{2}\right] ^{s}}\left( \nu
_{1,\tilde{m}}\widetilde{\mathbf{T}}^{-s}+\nu _{2,\tilde{m}}\widetilde{%
\mathbf{T}}^{s}\right) \left( \kappa _{J}^{L}\right) _{(\alpha _{1}\alpha
_{2}}\cdots \left( \kappa _{J}^{L}\right) _{\alpha _{2s-1}\alpha _{2s})}%
\text{ ,}
\end{eqnarray*}%
\begin{equation}
\end{equation}%
where $\nu _{1,\tilde{m}}=\sum_{m}\nu _{1,m,\tilde{m}}$ and $\nu _{2,\tilde{m}}=\sum_{m}\nu _{2,m,\tilde{m}}$, which we assume to be
finite and vanishing for $\tilde m=0$.

Thus, to summarize, if the Weyl zero-form depends on either $E$ or $J$, but not both, 
in the holomorphic gauge, then Weyl tensors in $L$-gauge become proportional to direct
products of $\kappa^L$'s, which means they are of generalized Petrov type D.

\paragraph{The case $\protect\theta \tilde{%
\protect\theta}\neq 0$.}

If both $\theta $ and $\tilde{\theta}$ are non-zero, \emph{i.e.} if both $E$ and$\ J$ are present
in the Weyl zero-form in the holomorphic gauge, then we can simplify the
analysis by substituting the explicit expressions provided in Appendices \ref%
{Sec gamma} and \ref{Sec coordinates} into the spin-$s$ Weyl tensor (\ref{Weyl tensor}%
) in $L$-gauge.

If $\mathbf{B}^{2}-\mathbf{C}^{2}=0$ \emph{i.e.} $m\theta =\pm \tilde{m}\tilde{\theta}$, then
the second set of terms in (\ref{Weyl tensor}) vanishes, and in the first set of terms $\mathbf{F}_{\alpha \beta }=\mathbf{B}\left[\left( \kappa _{E}^{L}\right) _{\alpha \beta }\pm \left(\kappa _{J}^{L}\right) _{\alpha \beta }\right]$.
This means that if $\theta/\tilde{\theta}$ is a rational number, and if furthermore we turn on only the terms with $m\theta =\pm \tilde{m}\tilde{\theta}$, then the Weyl tensors become proportional to direct products of $\left(\kappa _{E}^{L} \pm \kappa _{J}^{L} \right)$'s, \emph{i.e.} they are of generalized Petrov type D.%
\footnote{%
	Note, however, the matrix $\mathbf{K}_{\underline{\alpha \beta }}\equiv %
	\left[ \mathbf{B}E_{\underline{\alpha \beta }}+\mathbf{C}J_{\underline{%
			\alpha \beta }}\right] $ in this special case has determinant $\left(
	\mathbf{B}^{2}-\mathbf{C}^{2}\right) ^{2}=0$, which has 
	consequences for the twistor space connection; see Eq.\ \eqref{explicit detK}.}
However, for generic values of  $\theta/\tilde{\theta}$, (\ref{Weyl tensor}) is not of 
type D,%
\footnote
{See Appendix \ref{Sec Petrov type} for details on spin-2.}
though it is a sum of type-D tensors. 

\subsection{Asymptotic behaviour of the Weyl tensors}

By using the gamma matrix realization in Appendix 
\ref{Sec gamma} and the global coordinates in 
Appendix \ref{Sec coordinates}, we can investigate the asymptotic behaviour of the Weyl tensors. When $r\rightarrow \infty$, we have
\begin{equation}
\mathbf{F}^2|_{r\rightarrow\infty}\ =\ \mathbf{\breve F}^2|_{r\rightarrow\infty}
\ =\ \lambda^2 r^2\left[-\mathbf{B}^2+\mathbf{C}^2 {\rm sin}^2 (\vartheta)\right] \ ,
\end{equation}
Then the terms in (\ref{Weyl tensor}) of spin-$s$ Weyl tensor at large radius, by a simple power counting, scale as
\begin{equation}
\left\{\lambda^2 r^2\left[-\mathbf{B}^2+\mathbf{C}^2 {\rm sin}^2 (\vartheta)\right]\right\}^{-\frac{1}{2}(s+1)} \ ,
\end{equation}
and hence each term is either Kerr-like (when $\mathbf{B}^2\neq \mathbf{C}^2$) or 2-brane-like (when $\mathbf{B}^2 = \mathbf{C}^2$) in the asymptotic region. The Weyl tensor as the sum of these terms falls off as $\frac{1}{r^{s+1}}$, which is the regular boundary condition of asymptotically AdS$_{4}$ solutions.

\subsection{Zero-form charges} \label{Sec 0-form charges}

Although the separate spin-$s$ Weyl tensors blow up at the origin
of spacetime, the limit of the full Weyl zero-form remains well-defined as the 
symbol of an operator.
From this operator, it is possible to obtain higher spin gauge invariant 
quantities given by
\begin{equation}
\mathcal{I}_{2p}:=\int_{\mathcal{Z}_{4}} \text{Tr}^{\prime }\left\{
I\star \bar{I}\star \left[ \Phi \star \pi \left( \Phi \right) \right]
^{\star p}\right\} \text{ ,}\label{calI2p}
\end{equation}
which are referred to as zero-form charges \cite{Sezgin:2005pv}
and that are related to higher spin amplitudes 
\cite{Colombo:2010fu,Colombo:2012jx,Didenko:2012tv}.
On our exact solutions, \emph{i.e.} by substituting (\ref{AnsatzPhi}) and (%
\ref{ansatzF_YZ}), these charges are given by
\begin{equation}
\left. \mathcal{I}_{2p}\right\vert _{\text{on-solution}}:=\frac{1}{32}\sum
_{\substack{ \sigma , \\ m_{1},m_{2},\cdots ,m_{2p}, \\ \tilde{m}_{1},\tilde{m}_{2},\cdots ,\tilde{m}_{2p}}}\text{ }\mathbf{A}_{\tsum_{j=1}^{2p}
\left(-1\right) ^{j+1}m_{j},\tsum_{j=1}^{2p}\tilde{m}_{j}}\tprod_{j=1}^{2p}\nu _{\sigma ,m_{j},\tilde{m}_{j}}\text{ ,}
\end{equation}%
where%
\begin{equation}
\mathbf{A}_{m,\tilde{m}}:=\frac{\left[ \text{sech}(
m\theta ) \ \text{sech}( \tilde{m}\tilde{\theta}%
) \right] ^{2}}{1-\left[ \text{tanh}( m\theta ) \ 
\text{tanh}(  \tilde{m} \tilde{\theta}) \right] ^{2}}\text{
.}
\end{equation}%
The simplest case is $p=1$:%
\begin{equation}
\left. \mathcal{I}_{2}\right\vert _{\text{on-solution}}=\frac{1}{32}\sum
_{\substack{ \sigma ,m,\tilde{m},n,\tilde{n}}}\text{ }\mathbf{A}_{m-n,\tilde{%
m}+\tilde{n}}\nu _{\sigma ,m,\tilde{m}}\nu _{\sigma ,n,\tilde{n}}\text{ .}
\end{equation}
In \cite{Boulanger:2015kfa}, this zero-form charge has been proposed to
be one of the contributions to the effective action for 
higher spin gravity in asymptotically anti-de Sitter spacetimes.
As noted at the end of Section \ref{Sec biaxial sym}, the resulting 
contribution to the free energy functional is not positive
definite.
%

\section{Twistor space connection}\label{Sec analyticity}

In this section, we first compute the twistor space connection
$V^{(L)}_{\underline{\alpha}}$, and show in special cases that it admits a 
regular power series expansion on ${\cal Z}_4$ around 
$Z^{\underline{\alpha}}=0$ over finite regions of spacetime 
provided that the group algebra 
$\Comp[\mathbb Z\times\mathbb Z]$ is truncated
down to a non-unital subalgebra.
We then demonstrate the existence of the linearized 
gauge function $H^{(1)}$ taking the linearized twistor 
space connection to Vasiliev gauge in a special case.

\subsection{Generating function for twistor space connection in $L$-gauge
\label{Sec one-form analyticity}}

In order to facilitate the analysis, we write
\begin{eqnarray}
V_{\alpha }^{(L)} &=&L^{-1}\star V_{\alpha }^{\prime }\star L  \notag \\
&=&\sum_{\sigma ,m,\tilde{m}}L^{-1}\star T_{m,\tilde{m}}\star \Pi _{\sigma
}\star L\star \left( a_{\sigma ,m,\tilde{m}}\right) _{\alpha }\nonumber\\
&=&\sum_{\sigma ,m,\tilde{m}} 
\oint_{0}\frac{d\zeta }{2\pi i\zeta ^{m+1}}\oint_{0}\frac{d\tilde{\zeta}}{2\pi i\tilde{\zeta}^{\tilde m+1}}
\frac{1}{2}\left(\mathring V_{0;\sigma,m,\tilde m}^{(L)}
+\sigma \mathring V_{1;\sigma,m,\tilde m}^{(L)}\right)_\a
\ ,\label{VL}
\end{eqnarray}
in terms of the generating functions ($n=0,1$)
\begin{equation}
\left(\mathring V_{n;\sigma,m,\tilde m}^{(L)}\right)_\a := 
2i\frac{\partial}{\partial \rho^\a}
\int_{-1}^1\frac{d\tau  j_\s(\tau )
 }{\left( \tau +1\right) ^{2}}
T^L_{m,\tilde{m}} \star (\kappa _{y}\bar{\kappa}_{\bar{y}})^n \star 
\left\{ \text{exp}\left[\varsigma c\left( \tau \right) U^{\beta \gamma }z_{\beta
}z_{\gamma }+\rho ^{\beta }z_{\beta }\right] \right\} 
_{\rho=0}
\text{ ,} \qquad \label{analy2}
\end{equation}
where $\rho^\a$ is an auxiliary commuting spinor, and we denote $j_\s(\tau ) \equiv j_{\sigma }\left( \varsigma \mathring{\mu}_{\sigma };\tau \right)$.
Thus, if these two integrals are finite for bounded $\mathring{\mu}_{\sigma }$
and finite $\rho^\a$, then $V_{\alpha }$ is real-analytic in ${\cal Y}_4\times {\cal Z}_4$.

\subsection{Singular twistor space connection in $L$-gauge from $T_{0,0}$ \label{Sec truncation}}

From the discussion in Section \ref{Singular Integrand} and 
the fact that $T^L_{0,0}=1$, it follows that \eqref{VL} 
contains a term given by $(a_{\sigma,0,0})_\a$, which is 
not real-analytic in ${\cal Z}_4$. 
Thus, real-analyticity of $V_\a^{(L)}$ in ${\cal Z}_4$ requires 
\be (a_{\sigma,0,0})_\a=0\ .\ee
This can be achieved by a consistent truncation of the 
Ansatz (\ref{ansatzF_YZ})--(\ref{ansatzVbar_YZ}) by taking
${\cal A}_{E,J}$ to be a semigroup without 
identity.\footnote{Removing the unity from the presentation of 
${\cal A}$ also removes a singularity from $\Phi^{(L)}$.}

If $\theta \tilde{\theta} \neq 0$ this can be achieved by taking
\be \nu _{\sigma ,m,\tilde{m}}=(a_{\sigma,m,\tilde{m}})_\a=0\quad \mbox{\ for \ \ \  $m\leqslant 0$ \ and/or\ \
$\tilde m\leqslant 0$ .}\ee
In other words, in the original Ansatz we sum over $m,\tilde m\in\mathbb{Z}$, but due to the requirement of real-analyticity, we instead sum over only positive $m$ and/or positive $\tilde{m}$. Furthermore, as can be seen from the table (\ref{realcond-nu-sigma}), for compatibility with the reality condition, along with the truncation we must set $\theta $ and/or $\tilde{\theta}$ to be real.

If $\theta= 0$ (or $\tilde{\theta}=0$) then we need to restrict $\tilde{m}%
\in \mathbb{Z}^{+}$, $\tilde{\theta} \in \mathbb{R}$ (or $m\in \mathbb{Z}^{+}$, $\theta \in \mathbb{R}$).

$\theta $ and $\tilde{\theta}$ cannot be both zero.

To summarize, in the following table, we give notations to the consistent truncations, and ``$\times $'' means that the situation either includes the unity or is inconsistent with the reality condition.
\be
\begin{tabular}{cc|ccccc}
	&  & $\theta \in \mathbb{R}\backslash \{0\}$ & \multicolumn{1}{c}{$\theta
		\in \mathbb{R}\backslash \{0\}$} & $\theta \in i\mathbb{R}\backslash \{0\}$
	& \multicolumn{1}{c}{$\theta \in i\mathbb{R}\backslash \{0\}$} & $\ \ \
	\theta =0\ \ \ $ \\ 
	&  & $m\in \mathbb{Z}$ & $m\in \mathbb{Z}^{+}$ & $m\in \mathbb{Z}$ & $m\in 
	\mathbb{Z}^{+}$ &  \\ \hline
	$\tilde{\theta}\in \mathbb{R}\backslash \{0\}$ & $\tilde{m}\in \mathbb{Z}$ & 
	$\times $ & $\mathcal{A}_{+,\pm }$ & $\times $ & $\times $ & $\times $ \\ 
	$\tilde{\theta}\in \mathbb{R}\backslash \{0\}$ & $\tilde{m}\in \mathbb{Z}^{+}
	$ & $\mathcal{A}_{\pm ,+}$ & $\mathcal{A}_{+,+}$ & $\mathcal{A}_{\pm i,+}$ & 
	$\times $ & $\mathcal{A}_{0,+}$ \\ 
	$\tilde{\theta}\in i\mathbb{R}\backslash \{0\}$ & $\tilde{m}\in \mathbb{Z}$
	& $\times $ & $\mathcal{A}_{+,\pm i}$ & $\times $ & $\times $ & $\times $ \\ 
	$\tilde{\theta}\in i\mathbb{R}\backslash \{0\}$ & $\tilde{m}\in \mathbb{Z}%
	^{+}$ & $\times $ & $\times $ & $\times $ & $\times $ & $\times $ \\ 
	$\tilde{\theta}=0$ &  & $\times $ & $\mathcal{A}_{+,0}$ & $\times $ & $%
	\times $ & $\times $%
\end{tabular}
\label{TableTruncation}
\ee

\subsection{Regularity of twistor space connection in $L$-gauge 
for non-unital ${\cal A}_{E,J}$}

Under the assumption that ${\cal A}_{E,J}$ does not contain the unity,
we proceed by investigating (\ref{analy2}).
From (\ref{Tkappa}) it follows that
\begin{equation}
T_{m,\tilde{m}}\star \kappa _{y}\bar{\kappa}_{\bar{y}}=\frac{\mathbf{A}}{ 
\sqrt{\text{det}\left( \mathbf{K}\right) }}\text{exp}\left\{ -\frac{1}{2}%
\left( \mathbf{K}^{-1}\right) ^{\underline{\alpha \beta }}Y_{\underline{
\alpha }}Y_{\underline{\beta }}\right\} \text{ ,}
\end{equation}
where 
\be
\left( \mathbf{K}^{-1}\right) ^{\underline{\alpha \beta }} =\frac{1}{ 
\mathbf{B}^{2}-\mathbf{C}^{2}}\left[ \mathbf{B}E^{\underline{\alpha \beta }}-
\mathbf{C}J^{\underline{\alpha \beta }}\right] \text{ ,} \qquad
\text{det}\left( \mathbf{K}\right) =\left( \mathbf{B}^{2}-\mathbf{C}
^{2}\right) ^{2}\text{ .}  \label{explicit detK}
\ee
Thus the case of $n=1$ is equivalent to the case of $n=0$ 
by replacing $\mathbf{B}$ and $\mathbf{C}$ with 
$\frac{\mathbf{B}}{\mathbf{B}^{2}-\mathbf{C}^{2}}$ and $\frac{-\mathbf{%
C }}{\mathbf{B}^{2}-\mathbf{C}^{2}}$, respectively,
and multiplying an overall factor $\frac{1}{\sqrt{\left(\mathbf{B}^{2}-\mathbf{C}^{2}\right)^2}}$, 
which is possible provided that $\mathbf{B}
^{2}\neq \mathbf{C}^{2}$ \emph{i.e.} if $m\theta \neq \pm \tilde{m}\tilde{\theta}$
for all allowed values of $m$ and $\tilde{m}$ (in the non-unital case).
This can be achieved by a suitable choice of $\theta $ and $\tilde{\theta}$.

For $n=0$ we performing the $\star $-products between 
$T_{m,\tilde{m}}^L$ and the $Z$-dependent exponential
in (\ref{analy2}),
which yields 
\begin{flalign}
&\left(\mathring V_{0;m,\tilde m}^{(L)}\right)_\a \notag\\
=& \ 2i\int_{-1}^1 \frac{d\tau j_{\sigma }\left( \tau \right) }{\left( \tau +1\right) ^{2}} \frac{ 
	\mathbf{A}}{\sqrt{\mathbf{F}^{2} \mathbf{M}^2(\tau)}}
\text{exp}\left\{ -iz^{\beta }y_{\beta }- 
\frac{1}{2}\bar{y}^{\dot{\alpha}}\mathbf{H}_{\dot{\alpha}\dot{\beta}}\bar{y}^{\dot{
		\beta}}+\frac{\mathbf{F}^{\gamma \delta }}{2\mathbf{F}^{2}}\left( iz_{\gamma
}+\mathbf{G}_{\gamma \dot{\alpha}}\bar{y}^{\dot{\alpha}}\right) \left(
iz_{\delta }+\mathbf{G}_{\delta \dot{\beta}}\bar{y}^{\dot{\beta}}\right)
\right\} \notag\\
&\frac{\partial }{\partial \rho ^{\alpha }}\left\{ 
\text{exp}\left\{ \frac{\mathbf{M}_{\beta \gamma }(\tau)}{2
\mathbf{M}^2(\tau)} \left[ iy^{\beta }-\rho ^{\beta }+\frac{i\mathbf{F}^{\beta
\delta }}{\mathbf{F}^{2}}\left( iz_{\delta }+\mathbf{G}_{\delta \dot{\alpha
}}\bar{y}^{\dot{ \alpha}}\right) \right] \left[ iy^{\gamma }-\rho ^{\gamma }+
\frac{i\mathbf{F}^{\gamma \xi }}{\mathbf{F}^{2}}\left( iz_{\xi }+\mathbf{G}	_{\xi \dot{\beta}} \bar{y}^{\dot{\beta}}\right) \right] \right\} \right\}	_{\rho =0}\notag\\[5pt]
=& \ 2i\int_{-1}^1 
\frac{ d\tau j_{\sigma }\left( \tau \right) }{\left( \tau +1\right) ^{2}}\frac{ 
-\mathbf{A} \mathbf{M}_{\alpha \beta }(\tau)}{\sqrt{\mathbf{F}^{2}\mathbf{M}^2(\tau)} 
\mathbf{M}^2(\tau)}\left[ iy^{\beta }+\frac{i\mathbf{F}^{\beta \gamma }}
{\mathbf{F}^{2}}\left( iz_{\gamma }+\mathbf{G}_{\gamma \dot{\alpha}}\bar{y}^{\dot{
\alpha}}\right) \right]  \notag \\
&\ \ \ \ \ \ \ \ \ \ \ 
\text{exp}\left\{ -iz^{\beta }y_{\beta }-\frac{1}{2}\bar{y}^{\dot{\alpha}
}\mathbf{H}_{\dot{\alpha}\dot{\beta}}\bar{y}^{\dot{\beta}}+\frac{\mathbf{F}^{\gamma
\delta }}{2\mathbf{F}^{2}}\left( iz_{\gamma }+\mathbf{G}_{\gamma \dot{\alpha}
}\bar{y}^{\dot{\alpha}}\right) \left( iz_{\delta }+\mathbf{G}_{\delta \dot{
\beta}}\bar{y}^{\dot{\beta}}\right) \right.  \notag \\
&\ \ \ \ \ \ \ \ \ \ \ \ \ \  \left. \ \ +\frac{\mathbf{M}_{\beta \gamma }(\tau)}{2\mathbf{M}^2(\tau)}\left[
iy^{\beta }+\frac{i\mathbf{F}^{\beta \delta }}{\mathbf{F}^{2}}\left(
iz_{\delta }+\mathbf{G}_{\delta \dot{\alpha}}\bar{y}^{\dot{\alpha}}\right) 
\right] \left[ iy^{\gamma }+\frac{i\mathbf{F}^{\gamma \xi }}{\mathbf{F}^{2}}
\left( iz_{\xi }+\mathbf{G}_{\xi \dot{\beta}}\bar{y}^{\dot{\beta}}\right) 
\right] \right\} \text{ ,}  \label{analy1convert}
\end{flalign}
where 
\begin{equation}
\mathbf{M}_{\alpha \beta }(\tau)\equiv  \frac{\mathbf{F}_{\alpha \beta }}{ 
\mathbf{F}^{2}}-2\varsigma c\left( \tau \right) U_{\alpha \beta } \text{ ,}
\end{equation}
The integrand has potential divergencies at 
$\tau=0$, $\tau=-1$ and any value for $\tau$
where $\mathbf{F}^{2}$ or $\mathbf{M}^{2}(\tau)$ vanishes.
As analysed in Section \ref{Singular Integrand}, the
potential divergencies in $j_{\sigma }\left( \tau \right) $ 
at $\tau=0$ do not spoil the convergence of the integral provided
that the $\nu$- and $b_n$-parameters are sufficiently 
small. 
Furthermore, since $\mathbf{M}_{\alpha \beta }(\tau)\sim \left( \tau +1\right)
^{-1}$ as $(\tau+1)\rightarrow 0$, it follows that both the prefactor and
the exponent are bounded at $\tau=-1$.

To facilitate the investigation of $\mathbf{F}^{2}$ and
$\mathbf{M}^{2}(\tau)$, which are thus functions of 
$m\theta$, $\tilde{m}\tilde{\theta}$, ${\cal X}_4$ and $\tau$, 
we use the gamma matrix realization in Appendix 
\ref{Sec gamma} and the coordinates for $L$ in 
Appendix \ref{Sec coordinates}. 
We have not succeeded in a complete analysis,
but we have been able to cover a few important
special cases as follows:

\paragraph{The case $\mathcal{A}_{+,0}$.}

In this case, we have $\theta \in \mathbb{R}\backslash \{0\}$,  $\tilde{\protect\theta}=0$, $m\in \mathbb{Z}^{+}$, and hence $\mathbf{A}={\rm sech}^2 m\theta$, 
$\mathbf{B}={\rm \tanh}\, m\theta$, $\mathbf{C}=0$. 
Using the explicit matrices and spherical
coordinates defined in the appendices, we obtain 
\begin{eqnarray}
\mathbf{F}^{2} &=&-\mathbf{B}^{2}\lambda ^{2}r^{2}\text{ ,} \\
\mathbf{M}^{2}(\tau)&=&-\frac{\left[\varsigma c\left( \tau \right) \mathbf{B}\lambda
r+e^{i\vartheta }\right] \left[ \varsigma  c\left( \tau \right) \mathbf{B}\lambda
r+e^{-i\vartheta }\right] }{\mathbf{B}^{2}\lambda ^{2}r^{2}}\text{ .}
\end{eqnarray}
From $m\theta \neq 0$ it follows that 
$\mathbf{B}\neq 0$, and hence $\mathbf{F}^{2}$ 
does not vanish except at $r=0$. 
Moreover, since $c( \tau )$ is purely imaginary, 
the quantity $\varsigma c\left( \tau \right) \mathbf{B}\lambda r$
is purely imaginary as well.
Thus $\mathbf{M}^{2}(\tau)$ vanishes
iff
\be \vartheta =\frac{\pi }{2}\ ,\qquad  \tau\in \left\{ -\frac{1+ \mathbf{B}\lambda
r}{ 1-\mathbf{B}\lambda r} , - \frac{1- \mathbf{B}\lambda
r}{ 1+\mathbf{B}\lambda r}\right\}\ .
\label{equatorialsing}
\ee
Thus, in this case the twistor space connection is 
real-analytic everywhere away from the equatorial plane in the spherical coordinates.\footnote{%
On the equatorial plane, for a certain value of $\tau$ between the integration limits, zero-denominators appear in the integrand of \eqref{analy1convert} both on the exponent and in the factor in the front. We leave the consequence of this for future work.
}

\paragraph{The case $\tilde{\protect\theta}\neq 0$.}

When $\tilde{\theta}\neq 0$, we resort to case-by-case investigation. We
will only show two examples below.

For example, if we consider the region of small $r$, \emph{i.e.} a small spatial
sphere around the origin point, we have 
\begin{eqnarray}
\mathbf{F}^{2} &=&\mathbf{C}^{2}-2i\mathbf{BC}\lambda r\ \text{cos}\left(
\vartheta \right) +O\left( r^{2}\right) \text{ ,} \\
\mathbf{M}^{2}(\tau) &=&-\mathbf{C}^{-2}\left[\varsigma c\left( \tau \right) \mathbf{C}-i%
\right] ^{2}-2\mathbf{BC}^{-3}\left[\varsigma c\left( \tau \right) \mathbf{C}-i\right]
\lambda r\ \text{cos}\left( \vartheta \right) +O\left( r^{2}\right) \text{ .}
\end{eqnarray}
A valid choice of the parameters is $\theta \in \mathbb{R}\backslash \{0\}$, $%
\tilde{\theta}\in i\mathbb{R}\backslash \{0\}$ and $m\in \mathbb{Z}^{+}$, \emph{i.e.} the truncation $\mathcal{A}_{+,\pm i}$.
With this choice, we have $\mathbf{B}\in \mathbb{R}\backslash \{0\}$, $%
\mathbf{C}\in i\mathbb{R}$. 
Then for $\mathbf{C}\neq 0$ \emph{i.e.} $\tilde{m}\neq
0$, both first leading terms of $\mathbf{F}^{2} $\ and $\mathbf{M}^{2}(\tau) $ are
non-zero. 
For $\tilde{m}=0$, the discussion is the same as the above $\tilde\theta=0$ case. 
To summarize, the one-form field in this case is 
real-analytic in the small sphere except on the equatorial plane.

For another example, we consider the region of small $\vartheta $, \emph{i.e.} a
narrow cone around the axis of symmetry $\vartheta =0$, we have
\begin{eqnarray}
\mathbf{F}^{2}  &=&\left( \mathbf{C}-i\mathbf{B}\lambda r\right) ^{2}+O\left(
\vartheta^{2}\right) \text{ ,} \\
\mathbf{M}^2(\tau)&=&\left[ \frac{1}{\mathbf{C}-i\mathbf{B}\lambda r}-i\varsigma c\left(
\tau \right) \right] ^{2}+O\left( \vartheta ^{2}\right) \text{ .}
\end{eqnarray}
A valid choice of the parameters is $\theta \in \mathbb{R}\backslash \{0\}$, $
\tilde{\theta}\in \mathbb{R}\backslash \{0\}$ and $m\in \mathbb{Z}^{+}$ \emph{i.e.} the truncation $\mathcal{A}_{+,\pm }$.
With this choice, we have $\mathbf{B}\in \mathbb{R}\backslash \{0\}$, $
\mathbf{C}\in \mathbb{R}$. 
Then for $r\neq 0$ we have $\mathbf{C}-i\mathbf{B}%
\lambda r\notin \mathbb{R}$, and thus, with $i\varsigma c\left( \tau \right)\in 
\mathbb{R}$, both first leading terms of $\mathbf{F}^{2} $\ and 
$\mathbf{M}^2(\tau)$ are non-zero. 
The one-form field in this case is real-analytic in the
narrow cone around the axis $\vartheta =0$ excluding the origin point.

\subsection{Linearized twistor space connection in Vasiliev gauge}

Finally, let us check in a special case that it is indeed possible to bring the 
linearized twistor space connection to Vasiliev gauge
by means of a linearized gauge transformation, as described
in Section \ref{Sec Gauge function}, \emph{viz.}
\begin{equation}
V_{\alpha }^{(G)(1)}=V_{\alpha }^{(L)(1)}+\partial _{\alpha }H^{(1)}\text{ ,}
\end{equation}
where $H^{(1)}$ is formally given by
\begin{equation}
H^{(1)}=H^{(1)}\vert_{Z=0}-\frac{1}{z^{\beta }\partial _{\beta }}\left( z^{\alpha }V_{\alpha
}^{(L)(1)}\right) \text{ .}
\end{equation}
Note that, as explained around Eq.\ (\ref{equatorialsing}), 
in the case $\tilde{\theta}=0$ with the truncation $\mathcal{A}%
_{+,0}$, the regularity of the twistor space
connection at $\vartheta =\frac{\pi }{2}$ has not yet been verified in the $L$-gauge. However, we
expect that this problem would not exist in Vasiliev gauge.

To perform the check, we set $\vartheta =\frac{\pi }{2}$, $t=\phi=0$, $y^{\a}=\bar{y}^{\dot\a}=0$ and $\lambda =1$.
The resulting expression of the generating function for the $L$-gauge twistor 
space connection reads
\begin{equation}
\left. \left( \mathring{V}_{n=0;\sigma ,m}^{(L)}\right) _{\alpha
}\right\vert _{\lambda =1;\ \vartheta =\frac{\pi }{2},\ t=\phi =0;\ y=\bar{y}%
=0}=-2i \mathbf{A}\int_{-1}^{1}\frac{d\tau j_{\sigma }\left( \tau \right) }{\left( \tau
+1\right) ^{2}} P_{\alpha \beta }z^{\beta }\text{exp%
}\left\{ Q_{\alpha \beta }z^{\alpha }z^{\beta }\right\} \text{ ,}
\end{equation}%
where%
\begin{equation}
P_{\alpha \beta }=-\left[ 1+\mathbf{B}^{2}c^{2}\left( \tau \right) r^{2}%
\right] ^{-\frac{3}{2}}\left( 
\begin{array}{cc}
\mathbf{B}\varsigma c\left( \tau \right) r & -1 \\ 
1 & \mathbf{B}\varsigma c\left( \tau \right) r%
\end{array}%
\right) \text{ ,}
\end{equation}%
\begin{equation}
Q_{\alpha \beta }=-\frac{1}{2}\varsigma c\left( \tau \right) \left[ 1+\mathbf{B}%
^{2}c^{2}\left( \tau \right) r^{2}\right] ^{-1}\left( 
\begin{array}{cc}
-\mathbf{B}\varsigma c\left( \tau \right) r & 1 \\ 
1 & \mathbf{B}\varsigma c\left( \tau \right) r%
\end{array}%
\right) \text{ .}
\end{equation}
Going to Vasiliev gauge, we obtain 
\begin{eqnarray}
&&\left. \left( \mathring{V}_{n=0;\sigma ,m}^{(G)(1)}\right) _{\alpha
}\right\vert _{\lambda =1;\ \vartheta =\frac{\pi }{2},\ t=\phi =0;\ y=\bar{y}=0}\notag
\\
&=&-\frac{i\varsigma b_{0}\mathring{\nu}_{\sigma }}{2}\mathbf{A}z_{\alpha
}\int_{-1}^{1}\frac{d\tau }{\left( \tau +1\right) ^{2}}\left\{ Pe^{Q(z)}+%
\frac{e^{Q(z)}-1-Q(z)e^{Q(z)}}{Q^2(z)}z^{\alpha }S_{\alpha
}{}^{\beta }Q_{\beta \gamma }z^{\gamma }\right\} \text{,}
\label{VG10PQ}
\end{eqnarray}%
where%
\begin{equation}
Q(z):=Q_{\alpha \beta }z^{\alpha }z^{\beta }\text{ ,}
\end{equation}
and we have decomposed 
\begin{equation}
P_{\alpha \beta }=:P\varepsilon _{\alpha \beta }+S_{\alpha \beta }\text{ ,}\quad S_{[\alpha\beta]}=0\ ,
\end{equation}%
\emph{i.e.}%
\begin{equation}
P=\left[ 1+\mathbf{B}^{2}c^{2}\left( \tau \right) r^{2}\right] ^{-\frac{3}{2%
}}\text{ , \quad }S_{\alpha \beta }=-\left[ 1+\mathbf{B}^{2}c^{2}\left( \tau
\right) r^{2}\right] ^{-\frac{3}{2}}\mathbf{B}\varsigma c\left( \tau \right) r\left( 
\begin{array}{cc}
1 & 0 \\ 
0 & 1%
\end{array}%
\right) \text{ .}
\end{equation}%
The integrand of (\ref{VG10PQ}) can be converted into a total derivative of $%
\tau $:%
\begin{eqnarray}
&&\left. \left( \mathring{V}_{n=0;\sigma ,m}^{(G)(1)}\right) _{\alpha
}\right\vert _{\lambda =1;\ \vartheta =\frac{\pi }{2},\ t=\phi =0;\ y=\bar{y}%
=0}  \notag \\
&=&-\frac{\varsigma b_{0}\mathring{\nu}_{\sigma }}{2}\mathbf{A}z_{\alpha
}\int_{-1}^{1}d\tau \frac{\partial }{\partial \tau } \left\{
\frac{(e^{Q(z)}-1)(Q(z)P-z^{\alpha }S_{\alpha
		}{}^{\beta }Q_{\beta \gamma }z^{\gamma })}{\frac{\partial {Q^2(z)}}{\partial c}}\right\}\ ,
\end{eqnarray}
where, more explicitly, 
\begin{eqnarray}
&&\frac{(e^{Q(z)}-1)(Q(z)P-z^{\alpha }S_{\alpha
	}{}^{\beta }Q_{\beta \gamma }z^{\gamma })}{\frac{\partial {Q^2(z)}}{\partial c}}
\notag
\\
&=&
 -\frac{\sqrt{1+
		\mathbf{B}^{2}c^{2}\left( \tau \right) r^{2}}\left[ 1-\text{exp}\left( \frac{%
		\mathbf{B}c\left( \tau \right) ^{2}r\left( z^{1}+z^{2}\right) \left(
		z^{1}-z^{2}\right) -2\varsigma c\left( \tau \right) z^{1}z^{2}}{2\left[ 1+\mathbf{B}%
		^{2}c^{2}\left( \tau \right) r^{2}\right] }\right) \right] }{\mathbf{B}%
	\varsigma c\left( \tau \right) r\left( z^{1}+z^{2}\right) \left( z^{1}-z^{2}\right)
	-2z^{1}z^{2}}\ .
\end{eqnarray}
Thus, assigning the singularity in the interior of the integration domain 
its principal value, and using separate analytical continuations above and
below the singularity, one finds that it 
does not contribute, and hence
\be\left. \left( \mathring{V}_{n=0;\sigma ,m}^{(G)(1)}\right) _{\alpha
}\right\vert _{\lambda =1;\ \vartheta =\frac{\pi }{2},\ t=\phi =0;\ y=\bar{y}%
=0}  =\frac{b_{0}\mathring{\nu}_{\sigma }}{2}z_{\alpha }\mathbf{A}%
\frac{1-\text{exp}\left[ \frac{1}{2\mathbf{B}r}\left( z^{1}+z^{2}\right)
\left( z^{1}-z^{2}\right) \right] }{\left( z^{1}+z^{2}\right) \left(
z^{1}-z^{2}\right) }\text{ .}  \label{VG10}
\ee
We note that the limit $z^\alpha\rightarrow 0$ must be taken
after the integration over $\tau$ has been performed. 
This yields a well-defined limit, such that the twistor space 
connection is indeed real-analytic at $z^\a=0$.
If one instead takes the limit $z^\alpha\rightarrow 0$ under the integral, 
one ends up with a divergent integral; this divergence cannot, however, 
be interpreted as any pole or other singularity at $z^\a=0$.
Thus, the prescription that we use is the unique one 
leading to a sensible result.\footnote{This suggests 
that in more general perturbatively defined solutions to Vasiliev's 
equations obtained by repeated homotopy integration \cite{Vasiliev:1990en,Sezgin:2002ru} 
(see also \cite{Didenko:2014dwa}), the resulting auxiliary integrals should 
be performed prior to taking the limit $z^\alpha\rightarrow 0$; 
whether this prescription is actually unique and correct, remains to be 
investigated.
}
We note, however, that in the holomorphic gauge the corresponding operations 
commute, and, correspondingly, the twistor space connection is non-real-analytic 
at $z^\a=0$ in this gauge; see Section \ref{Singular Integrand}.

Similarly, we can also calculate for $n=1$:
\begin{equation}
\left. \left( \mathring{V}_{n=1;\sigma ,m}^{(G)(1)}\right) _{\alpha
}\right\vert _{\lambda =1;\ \vartheta =\frac{\pi }{2},\ t=\phi =0;\ y=\bar{y}%
=0}=\frac{ b_{0}\mathring{\nu}_{\sigma }}{2}z_{\alpha }\mathbf{A}%
\frac{1-\text{exp}\left[ \frac{\mathbf{B}}{2r}\left( z^{1}+z^{2}\right)
\left( z^{1}-z^{2}\right) \right] }{\mathbf{B}^{2}\left( z^{1}+z^{2}\right)
\left( z^{1}-z^{2}\right) }\text{ .}  \label{VG11}
\end{equation}
Finally, using the analog of Eq.\ (\ref{VL}) in Vasiliev gauge,
\emph{i.e.} replacing the label $(L)$ with $(G)(1)$ and 
substituting (\ref{VG10}) and (\ref{VG11}), we obtain
\begin{eqnarray}
&&\left. \mathring{V}_{\alpha }^{(G)(1)}\right\vert _{\lambda =1;\ \vartheta
=\frac{\pi }{2},\ t=\phi =0;\ y=\bar{y}=0}  \notag \\
&=&\frac{ b_{0}%
}{4}z_{\alpha }\sum_{\sigma ,m,\tilde{m}}\mathbf{A}\nu _{\sigma ,m,\tilde{m}}\left[ \frac{1-\text{exp}\left[ \frac{1}{2\mathbf{B%
}r}\left( z^{1}+z^{2}\right) \left( z^{1}-z^{2}\right) \right] }{\left(
z^{1}+z^{2}\right) \left( z^{1}-z^{2}\right) }\right.\notag
\\
&&\ \ \ \ \ \ \ \ \ \ \ \ \ \ \ \ \ \ \ \ \ \ \ \ \
\left.+\sigma \frac{1-\text{exp}%
\left[ \frac{\mathbf{B}}{2r}\left( z^{1}+z^{2}\right) \left(
z^{1}-z^{2}\right) \right] }{\mathbf{B}^{2}\left( z^{1}+z^{2}\right) \left(
z^{1}-z^{2}\right) }\right] \text{ .}
\label{VG1sum}
\end{eqnarray}

Indeed, starting in Vasiliev gauge, one can integrate
the equations of motion for the linearized twistor space
connection directly without factorizing the inner Klein
operator $\kappa$, with the result \cite{Vasiliev:1990en}
\begin{equation}
V_{\alpha }^{(G)(1)}=-\frac{ib_0}{2}z_{\alpha }\int_{0}^{1}d\tau \ \tau e^{iy^{\alpha
	}z_{\alpha }\tau }\left( \left. \Phi ^{(G)(1)}\right\vert _{y\rightarrow -z\tau
}\right) \text{ .}  \label{Vas integral}
\end{equation}
We note that, unlike the solution for the twistor space connection
obtained starting in the holomorphic gauge, which refers to a splitting 
of $z^\alpha$ into $z^\pm$ as in Section \ref{Sec general ansatz}, the above expression does 
not refer to any auxiliary spinor frame in $Z$ space.
From $\Phi^{(G)(1)}=\Phi^{(L)(1)}=\Phi^{(L)}$
it follows that \eqref{Vas integral} implies that
\begin{eqnarray}
&&\left. \mathring{V}_{\alpha }^{(G)(1)}\right\vert _{\lambda =1;\ \vartheta
=\frac{\pi }{2},\ t=\phi =0;\ y=\bar{y}=0}  \notag \\
&=&\frac{b_0}{2}z_{\alpha }\sum_{m,\tilde{m}}\mathbf{A}\left[ \nu _{1,m,\tilde{m}}\frac{%
1-\text{exp}\left[ \frac{1}{2\mathbf{B}r}\left( z^{1}+z^{2}\right) \left(
z^{1}-z^{2}\right) \right] }{\left( z^{1}+z^{2}\right) \left(
z^{1}-z^{2}\right) }\right.\notag
\\
&&\ \ \ \ \ \ \ \ \ \ \ \ \ \ \ 
\left. +\nu _{2,m,\tilde{m}}\frac{1-\text{exp}\left[ \frac{%
\mathbf{B}}{2r}\left( z^{1}+z^{2}\right) \left( z^{1}-z^{2}\right) \right] }{%
\mathbf{B}^{2}\left( z^{1}+z^{2}\right) \left( z^{1}-z^{2}\right) }\right] 
\text{ ,}
\end{eqnarray}
which one can readily identify with \eqref{VG1sum} upon using \eqref{nurelation}.

\section{Conclusion} \label{Sec Conclusions}

In this paper, we have given a new class of bi-axially symmetric 
solutions to Vasiliev's bosonic higher spin gravity model
using an Ansatz based on gauge functions and separation of
the dependence on the coordinates in twistor space.

This facilitates the construction of perturbatively exact 
solutions in a holomorphic gauge.
In this gauge, the spacetime connection vanishes, the Weyl zero-form 
is constant, \emph{i.e.} it depends only on the fiber coordinates,
while the twistor space connection depends on the twistor space
via a universal holomorphic function on $Z$-space with singularties 
at $z^\alpha=0$ that we have exhibited in the Weyl order in 
Section \ref{Singular Integrand}, and on the fiber coordinates via the zero-form 
integration constants.
We have then expanded the dependence on the fiber coordinates
in terms of the basis of a group algebra generated by the 
exponents of $\theta E$ and $\tilde\theta J$, where $E$ and 
$J$ are the generators the time-translational and rotational 
symmetries of the solutions.

We have then switched on the spacetime 
dependence using a vacuum gauge function $L$.
In the resulting gauge, which we refer to as $L$-gauge,
the spacetime connetion describes
an anti-de Sitter spacetime. The terms containing the unity of the internal algebra need to be removed, in order for the Weyl zero-form in $L$-gauge to be real-analytic on twistor space.
The resulting generalized spin-$s$ Weyl tensor, which
thus obeys the Bargmann-Wigner equation, is given 
by a sum of generalized Petrov type-D tensors that are asymptotically Kerr-like or 2-brane-like. 
For special values of the parameters, including the symmetry
enhanced cases, the spin-$s$ Weyl tensor is of generalized Petrov 
type D.

We have also shown that the twistor space connection in $L$-gauge,
provided that the group
algebra is truncated to a non-unital semigroup algebra 
as summarized in the table (\ref{TableTruncation}),
is real-analytic in finite spacetime regions for
a number of choices of parameters. 
In particular, in the spherically symmetric case, it is 
real-analytic everywhere away from the equatorial plane. The Ansatz 
introduces a fixed frame in $Z$-space that breaks the 
manifest spherical symmetry upon going to the normal order
in master fields with a $Z$-dependence.
At this plane, singularities may appear in auxiliary 
integrals, whose treatment requires analytical 
continuations in twistor space.
We have not spelled out the nature of the resulting 
contributions to the twistor space connection in $L$-gauge
in this work.

Finally, we have examined the problem of transforming
the master fields from the $L$-gauge to Vasiliev 
gauge at the linearized level. It is trivial in the 
case of the Weyl zero-form.
As for the linearized twistor space connection, we have
argued that the transformation exists at spacetime 
points where the connection is real-analytic in twistor 
space in $L$-gauge.
Among the remaining cases, we have focused on the potential
divergence at the equatorial plane in the spherically symmetric
case, which should be removed by the transformation, as
the twistor space connection in Vasiliev's gauge does not
refer to any fixed frame in $Z$-space.
Indeed we have verified that this is the case at the origin 
of the fiber space (\emph{i.e.} at $Y^{\underline \a}=0$), for general $z^\alpha$,
and consequently we have found agreement with the expression
for the twistor space connection in Vasiliev gauge 
obtained by direct integration.

Thus, more briefly, we have found families of exact bi-axially
symmetric solutions in the holomorphic and $L$-gauges, and 
we have verified that they can be brought to Vasiliev gauge
at the linearized level in a special case, leaving the more
general case as well as higher order perturbation for future
study.

We end our conclusions by commenting on future directions.
We have left a number of technical details unattended, that
we would like to examine more carefully.
Besides the issues related to real-analyticity of the 
linearized master fields in Vasiliev gauge, there is
the intriguing degenerate case $\mathbf B=\mathbf C$.
Moreover, by taking limits for $\theta$ and $\nu$-parameters
it is possible to make contact with the solutions found in \cite{Iazeolla:2011cb},
and more general Kerr-like extensions thereof by expanding the fiber subalgebra
using a combination of group algebra elements and endomorphisms
in Fock spaces.

More generally, we recall that the importance of Vasiliev's gauge at linearized level
is that, when combined with normal order, 
the linearized spacetine connection $W_\mu^{(G)(1)}$ 
has a $Y$-expansion at $Z=0$ in terms of unfolded
Fronsdal tensors and the initial data $H^{(1)}\vert_{Z=0}$
modulo gauge transformations.\footnote{If $H^{(1)}\vert_{Z=0}$
and the gauge parameters belong to the same class of functions
then $H^{(1)}\vert_{Z=0}$ describes pure gauge degrees 
of freedom. }
Exact solutions, however, are easier to find in Weyl order
using the gauge function method.
As far as we can see from the results here and elsewhere,
we expect there to be an agreement at the linearized level
between the holomorphic and Vasiliev gauges for a fairly 
large class of linearized zero-form initial data 
$\Phi^{\prime(1)}(Y)$, and it would be desirable to establish 
this correspondence more precisely, \emph{e.g.} by expanding
$\Phi^{\prime(1)}(Y)$ in terms of twistor space plane waves.

Turing to higher order perturbations, the next step is 
to compute the first subleading 
corrections to all master fields in Vasiliev gauge, and 
examine whether real-analyticity in twistor space 
for generic spacetime points constrains the initial 
data $\Phi^{\prime(n)}(Y)$ for the zero-form and 
$H^{(n)}\vert_{Z=0}$ for the gauge function, for $n=1,2$. 
This may lead to modified asymptotic boundary conditions 
in AdS$_4$ and corresponding corrections to the 
zero-form charges.  
In particular, as proposed in \cite{Boulanger:2015kfa},
the zero-form charge ${\cal I}_2$ is a contribution to the 
free energy functional.
The corresponding sesqui-linear form is not definite
on the representation space of the underlying higher spin 
symmetry algebra containing the initial data of our solutions.
There are additional contributions to the free energy, however,
that may lead to an interesting phase diagram.

The above analysis can also be performed for the closely 
related Kerr-like solutions outlined above.
More generally, one may consider relaxing the Vasiliev gauge 
as well as the smoothness conditions in twistor space, which 
may lead to more general noncommutive geometries with 
interesting properties.

\paragraph{Acknowledgements\,}

We are thankful to V. Didenko and C. Iazeolla for useful discussions
and correspondence.
We have also benefited from conversations with C. Arias, R. Aros,
R. Bonezzi, N. Boulanger, K. Crysostomos, K. Morand, R. Olea, E. Sezgin, E. Skvortsov,
M. Taronna, A. Torres Gomez, M. Valenzuela and M. Vasiliev.
The work of P. S. is supported by Fondecyt Regular grant N$^{\rm o}$
1140296, Conicyt grant DPI 20140115 and UNAB internal grant DI-1382-16/R.
The work of Y. Y. is supported by the Chilean Fondecyt Postdoc
Project N$^{\rm o}$ 3150692.

\newpage \appendix

\section{The $\star $-exponent}

Let $Y^\alpha$, $\alpha=1,\dots, N$, be oscillator
variables obeying
\be [Y^\alpha,Y^\beta]_\star = 2iC^{\alpha\beta}\ ,\ee
where $N$ is even and $C^{\alpha\beta}$ is invertible.
Denote%
\begin{equation}
w=\frac{1}{4}K_{\alpha \beta }Y^{\alpha }\star Y^{\beta }\text{ ,}
\label{def_w}
\end{equation}%
where $K_{\alpha\beta}$ is a constant matrix obeying
\begin{equation}
K_{\alpha \beta }=K_{\beta \alpha }\text{ ,\qquad }
K_{\alpha \beta }K^{\beta \gamma }=\delta _{\alpha }{}^{\gamma }\text{ ,}
\end{equation}
where indices are raised and lowered using the conventions
$Y^\alpha=C^{\alpha\beta}Y_\beta$, $Y_\beta=Y^\beta C_{\beta\alpha}$,
and $
C^{\alpha \beta }C_{\alpha \gamma }=\delta _{\gamma }{}^{\beta }$.
The $\star $-exponent is defined by the Taylor series of exponential
function with $\star $-products replacing ordinary products. 
In what follows we will compute the symbol in Weyl order of the $\star $-exponent%
\begin{equation}
g\equiv e_{\star }^{-2tw}\text{ .}  \label{def_g}
\end{equation}
From (\ref{def_g}) we can derive%
\begin{equation}
w\star g=-\frac{1}{2}\frac{\partial g}{\partial t}\text{ ,}  \label{wg_dg}
\end{equation}%
and to proceed we will compute the symbol of $w\star g$.
To do so we use the  identity
\begin{equation}
Y_{\alpha }\star f\left( Y\right) =Y_{\alpha }f\left( Y\right) +i\frac{%
\partial }{\partial Y^{\alpha }}f\left( Y\right) \text{ .}
\end{equation}
Thus
\begin{equation}
Y^{\alpha }\star Y^{\beta }=Y^{\alpha }Y^{\beta }+iC^{\alpha \beta }\text{ .}
\end{equation}%
Hence (\ref{def_w}) can also be written as%
\begin{equation}
w=\frac{1}{4}K_{\alpha \beta }Y^{\alpha }Y^{\beta }\text{ .}
\end{equation}
Using this, we can show that %
\begin{equation}
w\star g=wg-\frac{N}{8}\frac{\partial g}{\partial w}-\frac{1}{4}w\frac{%
\partial ^{2}g}{\partial w^{2}}\text{ .}  \label{w_star_g}
\end{equation}

\begin{proof}
\begin{eqnarray*}
w\star g &=&\frac{1}{4}K^{\alpha \beta }Y_{\alpha }\star Y_{\beta }\star g \\
&=&\frac{1}{4}K^{\alpha \beta }Y_{\alpha }\star \left( Y_{\beta }g+i\frac{%
\partial g}{\partial Y^{\beta }}\right) \\
&=&\frac{1}{4}K^{\alpha \beta }\left( Y_{\alpha }Y_{\beta }g+iY_{\alpha }%
\frac{\partial g}{\partial Y^{\beta }}+i\frac{\partial \left( Y_{\beta
}g\right) }{\partial Y^{\alpha }}-\frac{\partial ^{2}g}{\partial Y^{\alpha
}\partial Y^{\beta }}\right) \\
&=&wg+\frac{i}{2}K^{\alpha \beta }Y_{\alpha }\frac{\partial g}{\partial
Y^{\beta }}-\frac{1}{4}K^{\alpha \beta }\frac{\partial ^{2}g}{\partial
Y^{\alpha }\partial Y^{\beta }}\text{ .}
\end{eqnarray*}%
The last two terms can be further converted:%
\begin{eqnarray*}
\frac{i}{2}K^{\alpha \beta }Y_{\alpha }\frac{\partial g}{\partial Y^{\beta }}
&=&\frac{i}{2}K^{\alpha \beta }Y_{\alpha }\frac{\partial g}{\partial w}\frac{%
\partial w}{\partial Y^{\beta }} \\
&=&\frac{i}{2}K^{\alpha \beta }Y_{\alpha }\left( \frac{1}{2}K_{\beta \gamma
}Y^{\gamma }\right) \frac{\partial g}{\partial w} \\
&=&\frac{i}{4}Y_{\alpha }Y^{\alpha }\frac{\partial g}{\partial w} \\
&=&0\text{ ,}
\end{eqnarray*}%
\begin{eqnarray*}
-\frac{1}{4}K^{\alpha \beta }\frac{\partial ^{2}g}{\partial Y^{\alpha
}\partial Y^{\beta }} &=&-\frac{1}{4}K^{\alpha \beta }\frac{\partial }{%
\partial Y^{\alpha }}\left( \frac{\partial g}{\partial w}\frac{\partial w}{%
\partial Y^{\beta }}\right) \\
&=&-\frac{1}{4}K^{\alpha \beta }\frac{\partial }{\partial Y^{\alpha }}\left(
\frac{1}{2}K_{\beta \gamma }Y^{\gamma }\frac{\partial g}{\partial w}\right)
\\
&=&-\frac{1}{8}\frac{\partial }{\partial Y^{\alpha }}\left( Y^{\alpha }\frac{%
\partial g}{\partial w}\right) \\
&=&-\frac{1}{8}\delta _{\alpha }{}^{\alpha }\frac{\partial g}{\partial w}-%
\frac{1}{8}Y^{\alpha }\frac{\partial ^{2}g}{\partial w^{2}}\frac{\partial w}{%
\partial Y^{\alpha }} \\
&=&-\frac{N}{8}\frac{\partial g}{\partial w}-\frac{1}{8}Y^{\alpha }\left(
\frac{1}{2}K_{\alpha \beta }Y^{\beta }\right) \frac{\partial ^{2}g}{\partial
w^{2}} \\
&=&-\frac{N}{8}\frac{\partial g}{\partial w}-\frac{1}{4}w\frac{\partial ^{2}g%
}{\partial w^{2}}\text{ .}
\end{eqnarray*}%
Thus (\ref{w_star_g}) is proven.
\end{proof}

By substituting (\ref{w_star_g}), (\ref{wg_dg}) can be converted to%
\begin{equation}
wg-\frac{N}{8}\frac{\partial g}{\partial w}-\frac{1}{4}w\frac{\partial ^{2}g%
}{\partial w^{2}}=-\frac{1}{2}\frac{\partial g}{\partial t}\text{ .}
\label{wg_no_star_dg}
\end{equation}%
This differential equation can be solved by substituting the Ansatz%
\begin{equation}
g=a\left( t\right) e^{b\left( t\right) w}\text{ ,}  \label{Ansatz_g}
\end{equation}
which gives%
\begin{equation}
a\left( t\right) we^{b\left( t\right) w}-\frac{N}{8}a\left( t\right) b\left(
t\right) e^{b\left( t\right) w}-\frac{1}{4}a\left( t\right) b^{2}\left(
t\right) we^{b\left( t\right) w}=-\frac{1}{2}a^{\prime }\left( t\right)
e^{b\left( t\right) w}-\frac{1}{2}a\left( t\right) b^{\prime }\left(
t\right) we^{b\left( t\right) w}\text{ ,}  \label{dg_ab}
\end{equation}%
and this equation requires the following set of ordinary differential equations
for $a\left( t\right) $
and $b\left( t\right) $ to be satisfied:%
\begin{eqnarray}
-\frac{1}{2}a^{\prime }\left( t\right) &=&-\frac{N}{8}a\left( t\right)
b\left( t\right) \text{ ,} \\
-\frac{1}{2}a\left( t\right) b^{\prime }\left( t\right) &=&a\left( t\right)
\left( 1-\frac{1}{4}b^{2}\left( t\right) \right) \text{ .}
\end{eqnarray}%
The general solution is given by %
\begin{eqnarray}
a\left( t\right) &=&C_{2}\left[ \text{sech}\left( t+C_{1}\right) \right] ^{%
\frac{N}{2}}\text{ ,}  \label{a(t)withConst} \\
b\left( t\right) &=&-2\text{tanh}\left( t+C_{1}\right) \text{ ,}
\label{b(t)withConst}
\end{eqnarray}%
where $C_{1}$ and $C_{2}$ are constants.
These are determined by requiring that (\ref{Ansatz_g})
and (\ref{def_g}) stand for the same solution of (\ref{wg_no_star_dg}). It
is obvious that
\begin{equation}
\left. g\right\vert _{t=0}=e_{\star }^{0w}=1\text{ ,}
\end{equation}%
and hence%
\begin{equation}
\left. \left( wg-\frac{N}{8}\frac{\partial g}{\partial w}-\frac{1}{4}w\frac{%
\partial ^{2}g}{\partial w^{2}}\right) \right\vert _{t=0}=\left. w\star
g\right\vert _{t=0}=w\text{ .}
\end{equation}%
Consequently we have%
\begin{eqnarray}
a\left( 0\right) e^{b\left( 0\right) w} &=&1\text{ ,} \\
a\left( 0\right) we^{b\left( 0\right) w}-\frac{N}{8}a\left( 0\right) b\left(
0\right) e^{b\left( 0\right) w}-\frac{1}{4}a\left( 0\right) b^{2}\left(
0\right) we^{b\left( 0\right) w} &=&w\text{ .}
\end{eqnarray}%
Therefore,%
\begin{equation}
a\left( 0\right) =1\text{ and }b\left( 0\right) =0\text{\ .}
\end{equation}%
By substituting them into (\ref{a(t)withConst}) and (\ref{b(t)withConst}) we
can determine that%
\begin{equation}
C_{1}=0\text{ and }C_{2}=1\text{ .}
\end{equation}%
Then we derive that%
\begin{eqnarray}
a\left( t\right) &=&\left[ \text{sech}\left( t\right) \right] ^{\frac{N}{2}}%
\text{ ,} \\
b\left( t\right) &=&-2\text{tanh}\left( t\right) \text{ .}
\end{eqnarray}%
In this way we conclude
\begin{equation}
e_{\star }^{-2tw}=g=\left[ \text{sech}\left( t\right) \right] ^{\frac{N}{2}%
}e^{-2\text{tanh}\left( t\right) w}\text{ .}
\label{star and ordinary exponent}
\end{equation}

\section{Van der Waerden symbols and gamma matrices
\label{Sec gamma}}

To simplify some of the calculations in this paper, one can use a set of
explicit matrix expressions of Pauli matrices and gamma matrices, which for
example is given in this appendix.

\subsection{Pauli matrices}

We define the $\sigma $-matrices with two lower spinor indices%
\begin{equation}
\left( \sigma _{0}\right) _{\alpha \dot{\alpha}}=\left(
\begin{array}{cc}
1 & 0 \\
0 & 1%
\end{array}%
\right) \text{\ , \ }\left( \sigma _{1}\right) _{\alpha \dot{\alpha}}=\left(
\begin{array}{cc}
0 & 1 \\
1 & 0%
\end{array}%
\right) \text{\ , \ }\left( \sigma _{2}\right) _{\alpha \dot{\alpha}}=\left(
\begin{array}{cc}
0 & -i \\
i & 0%
\end{array}%
\right) \text{\ , \ }\left( \sigma _{3}\right) _{\alpha \dot{\alpha}}=\left(
\begin{array}{cc}
1 & 0 \\
0 & -1%
\end{array}%
\right) \text{ ,}
\end{equation}%
where $\sigma _{0}$ is the identity matrix and $\sigma _{1,2,3}$ are the
usual Pauli matrices. We also define their complex conjugate:%
\begin{equation}
\left( \bar{\sigma}_{0}\right) _{\dot{\alpha}\alpha }=\left(
\begin{array}{cc}
1 & 0 \\
0 & 1%
\end{array}%
\right) \text{\ , \ }\left( \bar{\sigma}_{1}\right) _{\dot{\alpha}\alpha
}=\left(
\begin{array}{cc}
0 & 1 \\
1 & 0%
\end{array}%
\right) \text{\ , \ }\left( \bar{\sigma}_{2}\right) _{\dot{\alpha}\alpha
}=\left(
\begin{array}{cc}
0 & i \\
-i & 0%
\end{array}%
\right) \text{\ , \ }\left( \bar{\sigma}_{3}\right) _{\dot{\alpha}\alpha
}=\left(
\begin{array}{cc}
1 & 0 \\
0 & -1%
\end{array}%
\right) \text{ .}
\end{equation}%
Then obviously we have%
\begin{equation}
\left( \sigma _{a}\right) _{\alpha \dot{\alpha}}=\left( \bar{\sigma}%
_{a}\right) _{\dot{\alpha}\alpha }\text{ .}
\end{equation}%
Furthermore, we use%
\begin{equation}
\varepsilon ^{\alpha \beta }=\varepsilon _{\alpha \beta }=\varepsilon ^{\dot{%
\alpha}\dot{\beta}}=\varepsilon _{\dot{\alpha}\dot{\beta}}=\left(
\begin{array}{cc}
0 & 1 \\
-1 & 0%
\end{array}%
\right)  \label{eps explicit}
\end{equation}%
to raise or lower indices (by NW-SE rules).

We also define%
\begin{gather}
\left( \sigma _{ab}\right) _{\alpha \beta }=-\left( \sigma _{ba}\right)
_{\alpha \beta }=\left( \sigma _{\lbrack a}\right) _{\alpha }^{\ \ \dot{%
\gamma}}\left( \bar{\sigma}_{b]}\right) _{\dot{\gamma}\beta }\text{ ,} \\
\left( \bar{\sigma}_{ab}\right) _{\dot{\alpha}\dot{\beta}}=-\left( \bar{%
\sigma}_{ba}\right) _{\dot{\alpha}\dot{\beta}}=\left( \bar{\sigma}%
_{[a}\right) _{\dot{\alpha}}^{\ \ \gamma }\left( \sigma _{b]}\right)
_{\gamma \dot{\beta}}\text{ .}
\end{gather}%
To write them explicitly:
\begin{gather}
\left( \sigma _{01}\right) _{\alpha \beta }=\left(
\begin{array}{cc}
-1 & 0 \\
0 & 1%
\end{array}%
\right) \text{ \ , }\left( \sigma _{02}\right) _{\alpha \beta }=\left(
\begin{array}{cc}
i & 0 \\
0 & i%
\end{array}%
\right) \text{ \ , }\left( \sigma _{03}\right) _{\alpha \beta }=\left(
\begin{array}{cc}
0 & 1 \\
1 & 0%
\end{array}%
\right) \text{ \ ,}  \notag \\
\left( \sigma _{12}\right) _{\alpha \beta }=\left(
\begin{array}{cc}
0 & i \\
i & 0%
\end{array}%
\right) \text{ \ , }\left( \sigma _{13}\right) _{\alpha \beta }=\left(
\begin{array}{cc}
1 & 0 \\
0 & 1%
\end{array}%
\right) \text{ \ , }\left( \sigma _{23}\right) _{\alpha \beta }=\left(
\begin{array}{cc}
-i & 0 \\
0 & i%
\end{array}%
\right) \text{ \ ,}  \notag \\
\left( \bar{\sigma}_{01}\right) _{\dot{\alpha}\dot{\beta}}=\left(
\begin{array}{cc}
-1 & 0 \\
0 & 1%
\end{array}%
\right) \text{ \ , }\left( \bar{\sigma}_{02}\right) _{\dot{\alpha}\dot{\beta}%
}=\left(
\begin{array}{cc}
-i & 0 \\
0 & -i%
\end{array}%
\right) \text{ \ , }\left( \bar{\sigma}_{03}\right) _{\dot{\alpha}\dot{\beta}%
}=\left(
\begin{array}{cc}
0 & 1 \\
1 & 0%
\end{array}%
\right) \text{ \ ,}  \notag \\
\left( \bar{\sigma}_{12}\right) _{\dot{\alpha}\dot{\beta}}=\left(
\begin{array}{cc}
0 & -i \\
-i & 0%
\end{array}%
\right) \text{ \ , }\left( \bar{\sigma}_{13}\right) _{\dot{\alpha}\dot{\beta}%
}=\left(
\begin{array}{cc}
1 & 0 \\
0 & 1%
\end{array}%
\right) \text{ \ , }\left( \bar{\sigma}_{23}\right) _{\dot{\alpha}\dot{\beta}%
}=\left(
\begin{array}{cc}
i & 0 \\
0 & -i%
\end{array}%
\right) \text{ \ .}
\end{gather}%
As shown above, the pair of spinor indices are symmetric.

\subsection{Gamma matrices}

We construct the explicit expressions of gamma matrices in the following way:%
\begin{equation}
\left( \Gamma _{a}\right) _{\underline{\alpha }}^{\ \ \underline{\beta }%
}=\left(
\begin{array}{cc}
0 & \left( \sigma _{a}\right) _{\alpha }^{\ \ \dot{\beta}} \\
\left( \bar{\sigma}_{a}\right) _{\dot{\alpha}}^{\ \ \beta } & 0%
\end{array}%
\right) \text{ ,}
\end{equation}%
whose explicit expressions are%
\begin{gather}
\left( \Gamma _{0}\right) _{\underline{\alpha }}^{\ \ \underline{\beta }%
}=\left(
\begin{array}{cccc}
0 & 0 & 0 & -1 \\
0 & 0 & 1 & 0 \\
0 & -1 & 0 & 0 \\
1 & 0 & 0 & 0%
\end{array}%
\right) \text{ , }\left( \Gamma _{1}\right) _{\underline{\alpha }}^{\ \
\underline{\beta }}=\left(
\begin{array}{cccc}
0 & 0 & 1 & 0 \\
0 & 0 & 0 & -1 \\
1 & 0 & 0 & 0 \\
0 & -1 & 0 & 0%
\end{array}%
\right) \text{ ,}  \notag \\[5pt]
\text{ }\left( \Gamma _{2}\right) _{\underline{\alpha }}^{\ \ \underline{%
\beta }}=\left(
\begin{array}{cccc}
0 & 0 & -i & 0 \\
0 & 0 & 0 & -i \\
i & 0 & 0 & 0 \\
0 & i & 0 & 0%
\end{array}%
\right) \text{ , \ }\left( \Gamma _{3}\right) _{\underline{\alpha }}^{\ \
\underline{\beta }}=\left(
\begin{array}{cccc}
0 & 0 & 0 & -1 \\
0 & 0 & -1 & 0 \\
0 & -1 & 0 & 0 \\
-1 & 0 & 0 & 0%
\end{array}%
\right) \text{ ,}
\end{gather}%
One can check the property that%
\begin{equation}
\left( \Gamma _{(a}\right) _{\underline{\alpha }}^{\ \ \underline{\gamma }%
}\left( \Gamma _{b)}\right) _{\underline{\gamma }}^{\ \ \underline{\beta }%
}=\eta _{ab}\delta _{\underline{\alpha }}{}^{\underline{\beta }}\text{ ,}
\end{equation}%
where $\eta _{ab}=$ diag$\left\{ -1,1,1,1\right\} $.

We further use%
\begin{equation}
C^{\underline{\alpha \beta }}=\left(
\begin{array}{cc}
\varepsilon ^{\alpha \beta } & 0 \\
0 & \varepsilon ^{\dot{\alpha}\dot{\beta}}%
\end{array}%
\right) \text{ and }C_{\underline{\alpha \beta }}=\left(
\begin{array}{cc}
\varepsilon _{\alpha \beta } & 0 \\
0 & \varepsilon _{\dot{\alpha}\dot{\beta}}%
\end{array}%
\right) \text{ ,}
\end{equation}%
to raise or lower the spinor indices of gamma matrices (by NW-SE rules). For
example, by lowering the second spinor index, we get
\begin{equation}
\left( \Gamma _{a}\right) _{\underline{\alpha \beta }}=\left( \Gamma
_{a}\right) _{\underline{\alpha }}^{\ \ \underline{\gamma }}C_{\underline{%
\gamma \beta }}=\left(
\begin{array}{cc}
0 & \left( \sigma _{a}\right) _{\alpha \dot{\beta}} \\
\left( \bar{\sigma}_{a}\right) _{\dot{\alpha}\beta } & 0%
\end{array}%
\right) \text{ .}
\end{equation}%
One can check that in this way of construction, the pair spinor indices are
symmetric, \emph{i.e.} $\left( \Gamma _{a}\right) _{\underline{\alpha \beta }%
}=\left( \Gamma _{a}\right) _{\underline{\beta \alpha }}$.

We also define%
\begin{equation}
\left( \Gamma _{ab}\right) _{\underline{\alpha }}^{\ \ \underline{\beta }%
}=\left( \Gamma _{\lbrack a}\right) _{\underline{\alpha }}^{\ \ \underline{%
\gamma }}\left( \Gamma _{b]}\right) _{\underline{\gamma }}^{\ \ \underline{%
\beta }}\text{ .}
\end{equation}%
One can easily check that%
\begin{equation}
\left( \Gamma _{ab}\right) _{\underline{\alpha \beta }}=\left(
\begin{array}{cc}
\left( \sigma _{ab}\right) _{\alpha \beta } & 0 \\
0 & \left( \bar{\sigma}_{ab}\right) _{\dot{\alpha}\dot{\beta}}%
\end{array}%
\right) \text{ .}
\end{equation}%
In this way of construction, $\left( \Gamma _{ab}\right) _{\underline{\alpha
\beta }}=\left( \Gamma _{ab}\right) _{\underline{\beta \alpha }}$.

Now we define%
\begin{eqnarray}
E_{\underline{\alpha \beta }} &=&-\left( \Gamma _{0}\right) _{\underline{%
\alpha \beta }}=\left(
\begin{array}{cccc}
0 & 0 & -1 & 0 \\
0 & 0 & 0 & -1 \\
-1 & 0 & 0 & 0 \\
0 & -1 & 0 & 0%
\end{array}%
\right) \text{ ,}  \label{explicit E} \\[5pt]
J_{\underline{\alpha \beta }} &=&-\left( \Gamma _{12}\right) _{\underline{%
\alpha \beta }}=\left(
\begin{array}{cccc}
0 & -i & 0 & 0 \\
-i & 0 & 0 & 0 \\
0 & 0 & 0 & i \\
0 & 0 & i & 0%
\end{array}%
\right) \text{ .}  \label{explicit J}
\end{eqnarray}%
One can for instance check the properties (\ref{EJProp1})-(\ref{EJProp4})
using the above explicit matrix expressions.

\section{Spacetime gauge function\label{Sec coordinates}}

The four-dimensional anti-de Sitter spacetime, AdS$_{4}$,
of inverse radius $\lambda$, is the hyperbola
\begin{equation}
X^{A}X^{B}\eta _{AB}=-\lambda ^{-2}\,\ \text{,}
\end{equation}
in the five-dimensional space with coordinates $X^A$,
$A=\left\{ 0,1,2,3,0^{\prime }\right\} $ and flat
metric $\eta _{AB}=$diag$\left\{ -1,1,1,1,-1\right\} $.
A set of global coordinates
\begin{equation}\left( t,r,\vartheta ,\phi \right)\ ,\qquad
0\leqslant  \lambda t< 2\pi\ ,\quad
r\geqslant 0\ ,\quad 0\leqslant  \vartheta \leqslant  \pi\ ,\quad
 0\leqslant  \phi<2\pi\ , \end{equation}
can be introduced by taking
\begin{eqnarray}
X^{0} &=&-\sqrt{\lambda ^{-2}+r^{2}}\ \text{sin} \lambda t \text{
\ \ , \ \ \ }X^{0^{\prime }}=-\sqrt{\lambda ^{-2}+r^{2}}\ \text{cos}
\lambda t \text{ \ \ ,}  \notag \\
X^{1} &=&r\ \text{sin}\vartheta \ \text{cos}\phi \text{ \ \ , \ \ \ }%
X^{2}=r\ \text{sin}\vartheta \ \text{sin}\phi \text{ \ \ , \ \ \ }X^{3}=r\
\text{cos}\vartheta \text{ \ \ .}
\end{eqnarray}
The resulting induced metric is
\begin{equation}
ds^{2}=-\left( 1+\lambda ^{2}r^{2}\right) dt^{2}+\left( 1+\lambda
^{2}r^{2}\right) ^{-1}dr^{2}+r^{2}\left( d\vartheta ^{2}+\text{sin}%
^{2}\vartheta \ d\phi ^{2}\right) \text{ .}
\end{equation}
The stereographic coordinates
\begin{equation}
x^{\mu }\equiv \delta _{a}^{\mu }x^{a}=\frac{X^{a}}{1+|X^{0'}|}\ ,
\end{equation}
where $a=\left\{ 0,1,2,3\right\} $, $\eta_{ab} =\ $diag$\left\{ -,+,+,+\right\}$
and $x^{2}:=x^{a}x^{b}\eta_{ab}$, maps the two halves $X^{0'}>0$ and
$X^{0'}<0$ of AdS$_4$ to the region $-1<\lambda^2 x^2<1$ of $\Real^{3,1}$.
From the inverse relation given by
\begin{equation}X^{a}=\frac{2x^{a}}{1-\lambda ^{2}x^{2}}\ ,\quad X^{0'}=\pm \lambda^{-1}
\frac{1+\lambda^{2}x^2}{1-\lambda^{2}x^2}\ ,\end{equation}
it follows that $X^{0'}\rightarrow -X^{0'}$ corresponds to $x^a\rightarrow -(\lambda^2 x^2)^{-1}x^a$.
Thus, the extension of the stereographic coordinates $x^a$ to the
entire $\Real^{3,1}$ provides a global coordinate of AdS$_4$;
the boundary of AdS$_4$ is mapped to the hyperbola $\lambda^2 x^2=1$
in $\Real^{3,1}$.

The gauge function
\begin{equation}
L\left( x;y,\bar{y}\right) =\frac{2h}{1+h}\text{exp}\left( \frac{i\lambda }{%
1+h}x^{\alpha \dot{\alpha}}y_{\alpha }\bar{y}_{\dot{\alpha}}\right)\ ,
\qquad x_{\alpha \dot{\alpha}}:=x^{a}\left( \sigma
_{a}\right) _{\alpha \dot{\alpha}}\ ,\quad h:=\sqrt{1-\lambda
^{2}x^{2}}\text{ ,}
\end{equation}
which is defined in the region $\lambda^2 x^2<1$, leads to
\begin{equation}
U_{\mu }=L^{-1}\star \partial _{\mu }L=-\frac{i}{2}e_{\mu }^{\alpha \dot{%
\alpha}}y_{\alpha }\bar{y}_{\dot{\alpha}}-\frac{i}{4}\left( \omega _{\mu
}^{\alpha \beta }y_{\alpha }y_{\beta }+\bar{\omega}_{\mu }^{\dot{\alpha}\dot{%
\beta}}\bar{y}_{\dot{\alpha}}\bar{y}_{\dot{\beta}}\right) \text{ ,}
\end{equation}
where
\begin{eqnarray}
e_{\mu }^{\alpha \dot{\alpha}} &=&-\frac{\lambda \delta _{\mu }^{a}\left(
\sigma _{a}\right) ^{\alpha \dot{\alpha}}}{h^{2}}\text{\ ,} \\
\omega _{\mu }^{\alpha \beta } &=&-\frac{\lambda ^{2}\delta _{\mu
}^{a}x^{b}\left( \sigma _{ab}\right) ^{\alpha \beta }}{h^{2}}\text{\ , \ }%
\bar{\omega}_{\mu }^{\dot{\alpha}\dot{\beta}}=-\frac{\lambda ^{2}\delta
_{\mu }^{a}x^{b}\left( \bar{\sigma}_{ab}\right) ^{\dot{\alpha}\dot{\beta}}}{%
h^{2}}\text{ ,}
\end{eqnarray}%
are the vierbein and Lorentz connection
of AdS$_{4}$ in stereographic coordinates,
with flat indices converted to spinor
ones using van der Waerden symbols.
One also has
\begin{equation}
L^{-1}\star Y_{\underline{\alpha }}\star L=L_{\underline{\alpha }}{}^{%
\underline{\beta }}Y_{\underline{\beta }}\text{ ,}
\end{equation}%
with the matrix
\begin{equation}
L_{\underline{\alpha }}{}^{\underline{\beta }}=h^{-1}\left[
\begin{array}{cc}
\delta _{\alpha }{}^{\beta } & \lambda x_{\alpha }{}^{\dot{\beta}} \\
\lambda x_{\dot{\alpha}}{}^{\beta } & \delta _{\dot{\alpha}}{}^{\dot{\beta}}%
\end{array}%
\right] \text{ .}
\end{equation}
As an Sp(4;$\mathbb{R}$) group element, $L(x;y,\bar y)$
corresponds to the transvection in AdS$_{4}$ that
sends all the information of the classical solution
encoded at the origin of the stereographic
coordinate system to the point $x^\mu$.

\section{Determination of Petrov type of spin-2 Weyl tensor\label{Sec Petrov type}}

In this appendix, we briefly explain how to check (only for spin-2) the
Petrov type of a Weyl tensor by using the eigenvalue method. For more
details on this topic one can check \cite{Stephani:2003tm}.

The restricted Lorentz group SO$^{+}$(3,1,$\mathbb{R}$) is isomorphic to
SO(3,$\mathbb{C}$), and a Weyl tensor can be converted into its equivalent
form with SO(3,$\mathbb{C}$) indices. We can convert the Weyl tensor $%
C_{\alpha \beta \gamma \delta }$ with four symmetric SL(2;$\mathbb{C}$)
indices into an equivalent tensor $Q_{IJ}$ with two symmetric and traceless
SO(3,$\mathbb{C}$) indices, simply by using the Pauli matrices:%
\begin{equation}
Q_{IJ}=\left( \sigma _{I}\right) ^{\alpha \beta }\left( \sigma _{J}\right)
^{\gamma \delta }C_{\alpha \beta \gamma \delta }\text{ ,}
\end{equation}%
where $\left( \sigma _{I}\right) ^{\alpha \beta }=\varepsilon ^{\alpha
\alpha ^{\prime }}\left( \sigma _{I}\right) _{\alpha ^{\prime }}{}^{\beta }$
and we can explicitly choose%
\begin{equation}
\left( \sigma _{1}\right) _{\alpha }{}^{\beta }=\left(
\begin{array}{cc}
0 & 1 \\
1 & 0%
\end{array}%
\right) \text{\ , \ }\left( \sigma _{2}\right) _{\alpha }{}^{\beta }=\left(
\begin{array}{cc}
0 & -i \\
i & 0%
\end{array}%
\right) \text{\ , \ }\left( \sigma _{3}\right) _{\alpha }{}^{\beta }=\left(
\begin{array}{cc}
1 & 0 \\
0 & -1%
\end{array}%
\right) \text{ .}
\end{equation}%
The indices $I,J=1,2,3$ should be raised or lowered by the Kronecker delta,
so whether they are upper or lower indices does not make a difference.

If we treat $Q$ as a 3$\times $3 matrix, then observing its eigenvalues and
eigenvectors is sufficient for determining its Petrov type. Below we list
all Petrov types and their corresponding $Q$-matrix criteria:%
\begin{eqnarray}
&&%
\begin{tabular}{cc}
Petrov types & $Q$-matrix criteria \\
I & $\left[ Q-\lambda _{1}I\right] \left[ Q-\lambda _{2}I\right] \left[
Q-\lambda _{3}I\right] =0$ \\
D & $\left[ Q-\left( -\frac{1}{2}\lambda \right) I\right] \left[ Q-\lambda I%
\right] =0$ \\
II & $\left[ Q-\left( -\frac{1}{2}\lambda \right) I\right] ^{2}\left[
Q-\lambda I\right] =0$ \\
N & $Q^{2}=0$ \\
III & $Q^{3}=0$ \\
O & $Q=0$%
\end{tabular}
\notag \\
&&
\end{eqnarray}%
In the list, $\lambda _{1,2,3}$, $\lambda $ and $\left( -\frac{1}{2}\lambda
\right) $ are eigenvalues of $Q$, $\lambda _{1}+\lambda _{2}+\lambda _{3}=0$
and $I$ is the identity matrix. In particular, being Petrov type D means the
matrix $Q$ has three independent eigenvectors while two of them correspond
to equal eigenvalues.

Using the explicit matrices and coordinates provided in Appendices \ref{Sec gamma} and \ref{Sec coordinates}, for spin $s=2$ we can evaluate the Weyl
tensor (\ref{Weyl tensor}) at a given spacetime point with a chosen set of
parameters, and then we can evaluate the corresponding $Q$ matrix to check
its Petrov type. We have found that in general the $Q$ matrix has three
distinct eigenvalues (type I) and thus is not of type D, unless we choose
some special parameters or consider only some special spacetime locations.

\bibliography{biblio}
\bibliographystyle{utphys}

\end{document}